%% file: paper.tex
\documentclass[10pt,sigconf,letterpaper]{acmart}

\usepackage[english]{babel}
\usepackage{blindtext}

\renewcommand\footnotetextcopyrightpermission[1]{} 
\setcopyright{none}
\usepackage{placeins}

\acmConference[]{}

\acmPrice{}

\usepackage{tabularx}
\newcolumntype{L}[1]{>{\raggedright\arraybackslash}p{#1}}
\newcolumntype{C}[1]{>{\centering\arraybackslash}p{#1}}
\newcolumntype{R}[1]{>{\raggedleft\arraybackslash}p{#1}}

\usepackage{algorithm}
\usepackage[noend]{algorithmic}

\usepackage{subfig}
\usepackage{enumitem}
\setlist{nolistsep}
\usepackage{graphicx}
\usepackage{epstopdf}
\usepackage{grffile}
\usepackage{amssymb}
\usepackage{amsmath}
\usepackage{multicol,multirow}
\usepackage{rotating,url}
\usepackage{enumitem}
\usepackage{color}

\usepackage{xspace}
\usepackage{ifpdf}
\usepackage{mdwlist}
\usepackage{colortbl}
\usepackage{caption}
\usepackage{amssymb}
\usepackage{booktabs}
\usepackage{multirow}
\usepackage{enumitem}
\usepackage{xfrac}
\usepackage{afterpage}
\usepackage{pdflscape}
\usepackage{longtable}
\usepackage{hhline}
\usepackage{lscape}

\newlength{\oldtextfloatsep}
\newcommand{\ie}{i.e., \@}
\newcommand{\eg}{e.g., \@}

\newcommand{\etc}{etc. \@}

\usepackage{enumitem}
\usepackage{mdwlist}
\setlist{nolistsep}
\setlist[description]{noitemsep,topsep=0pt,parsep=0pt,partopsep=0pt,leftmargin=0pt}
\usepackage{amssymb}
\usepackage{pifont}

\usepackage{everypage}

\newcommand{\eat}[1]{}

\newcommand{\todo}[1]{\textcolor{red}{TODO: \emph{#1}}}

\definecolor{ballblue}{rgb}{0.13, 0.67, 0.8}
\newcommand{\cinote}[1]{\textcolor{ballblue}{CI-Note:[\emph{#1}]}}
\definecolor{fashionfuchsia}{rgb}{0.96, 0.0, 0.63}
\newcommand{\nlnote}[1]{\textcolor{fashionfuchsia}{\small \bf [NL: #1]}} 

\newcommand{\gsnote}[1]{\textcolor{blue}{\small \bf [GS: #1]}}

\long\def\comment#1{}

\newcommand{\Lpagenumber}{\ifdim\textwidth=\linewidth\else\bgroup
  \dimendef\margin=0 
  \ifodd\value{page}\margin=\oddsidemargin
  \else\margin=\evensidemargin
  \fi
  \raisebox{\dimexpr -\topmargin-\headheight-\headsep-0.5\linewidth}[0pt][0pt]{%
    \rlap{\hspace{\dimexpr \margin+\textheight+\footskip}%
    \llap{\rotatebox{90}{\thepage}}}}%
\egroup\fi}
\AddEverypageHook{\Lpagenumber}%

\begin{document}


\author{Costas Iordanou}
\affiliation{
  \institution{Max Planck Informatics}
}
  \email{iordanou@mpi-inf.mpg.de}

\author{Georgios Smaragdakis}
\affiliation{
  \institution{TU Berlin}
}
\email{georgios@inet.tu-berlin.de}

\author{Nikolaos Laoutaris}
\affiliation{
  \institution{IMDEA Networks Institute}
}
  \email{nikolaos.laoutaris@imdea.org}

\clubpenalty=10000 
\widowpenalty = 10000

\title[]{Who's Tracking Sensitive Domains?}
\subtitle{(and how can you tell a sensitive domain anyway?)}





\input{sections/abstract}

\maketitle
\input{sections/intro}

\input{sections/methodology}
\input{sections/classification}
\input{sections/trackers}

\input{sections/communication}
\input{sections/related}
\input{sections/conclusion}

%

%
\bibliographystyle{acm}
\bibliography{paper}
%


\end{document}

%% file: sections/abstract.tex
\begin{abstract}
We turn our attention to the \emph{elephant in the room} of data protection, which is none other than the
simple and obvious question: ``Who's tracking sensitive domains?''. Despite a fast-growing amount of work on more complex facets of
the interplay between privacy and the business models of the Web, the obvious question of who collects data on domains
where most people would prefer not be seen, has received rather limited attention. First, we develop a methodology for
automatically annotating websites that belong to a sensitive category, \eg as defined by the General Data
Protection Regulation (GDPR). Then, we extract the third party tracking services included directly, or via recursive
inclusions, by the above mentioned sites. Having analyzed around 30k sensitive domains, we show that such domains are
tracked, albeit less intensely than the mainstream ones. Looking in detail at the tracking services operating on them, we
find well known names, as well as some less known ones, including some specializing on specific sensitive categories.
\end{abstract}


%% file: sections/intro.tex
\section{Introduction}\label{sec:intro}


The public and scientific interest around data protection, privacy, and their relationship with new services and
business models on the Web is reaching an all-time-high. The intense debates and the pressing needs for evidence-based
policymaking have triggered lots of measurement work in the area~\cite{Ads-vs-Regular-Contents,
openWPM-englehardt2016census, WebTracking-over-years, Fruchter_2015, Walls_2015,
Balebako_2012, Pujol_2015,Mobile-Apps-NDSS2018,Narseo_2012,Binns2018,Leung_2016}.

Research in many cases has jumped directly to asking very general questions such as ``Who is
tracking?''~\cite{openWPM-englehardt2016census, Mayer_2012, Mobile-Apps-NDSS2018} and ``How is tracking
done?''~\cite{cookies-sync2019,Acar_2013,Starov_2016}, or proposing holistic solutions to privacy
challenges~\cite{PrivacyMeter,tracking-trackers}.  Apart from the technical difficulties related to such endeavours,
definitional, ethical, and other debates make these matters even more complex. E.g., to what extent is it justifiable to
collect and sell  end-user information in exchange for a free service? This question is inadvertently present
in any study of tracking on the Web, as well as in any proposal for stopping it or conducting it in a different manner.
Opinions on the above question vary, and this makes it difficult to reach a clear conclusion when analyzing the findings of general studies about tracking.

Yet, there exist some matters related to privacy in which most people agree. For example, most would not prefer to be tracked when visiting
domains involving sensitive topics such as religion, health and sexual orientation. This is so evident that it even
appears as an explicit clause in most data protection regulations, including 
the EU General Data Protection
Regulation (GDPR)~\cite{EU-GDPR} that considers as sensitive personal data any data \textit{``revealing racial or ethnic
origin, political opinions, religious or philosophical beliefs, or trade union membership, also genetic data, biometric
data for the purpose of uniquely identifying a natural person, data concerning health or data concerning a natural
persons sex life or sexual orientation''}. Other governments and administrations around the world, \eg in California (California
Consumer Privacy Act (CCPA)~\cite{CCPA}), Canada~\cite{Canada-privacy}, Israel~\cite{Israel-privacy},
Japan~\cite{JPIP}, and Australia~\cite{Australia-privacy}, are following similar
paths~\cite{Kalman-CACM-GDPR,CACM-GDPR-2018}. The public has taken notice and has started to make use of the new
regulations. Indeed, in the first seven months of GDPR there have been 95k filed complaints relating to
it~\cite{EU-dataprotection}; in France, this represented an increase of more than 60\%~\cite{Kalman-CACM-GDPR} compared
to the previous year.

In this paper, we investigate and report on the entities that track and collect data in web domains where most people
would rather prefer not to be seen by third parties. Being tracked on a cancer discussion forum, a gay dating site, or a news site
with non-mainstream political affinity, is at the core of some of the most fundamental anxieties that many people have
about their online privacy. Many people visit such sites in incognito mode. This can provide some privacy in some cases but it has been shown that tracking can be performed regardless~\cite{incognito-problems,private-browsing-2010,Private-browsing-2014}. In any case, our focus here is on answering whether such domains are being tracked. Quantifying the consequences of such tracking goes beyond the scope of this paper.   

\subsection{Challenges}


Answering the seemingly straightforward question posed by this paper is far from simple. As mentioned already, categories such as
health and sexual orientation, have been recorded on data protection laws, albeit at a high and abstract level. This makes sense, since such laws are meant to be interpreted by humans in courts of law. Thus, following a complaint, it will fall upon a data protection authority, or eventually a judge, to decide whether tracking people visiting a particular website violates any clause about sensitive data, a matter on which we do not have any opinion and consider outside the scope of our work.   

The objective of our work, however, is to answer at scale whether sensitive domains are tracked, and to do this we first
need to collect thousands of such domains. This cannot be done manually by picking one by one individual domains and,
therein, lies our first major challenge: ``how can we classify arbitrary domains programmatically as sensitive or not
and collect enough of them for a systematic analysis of tracking?''. The obvious approach of compiling a list of textual
descriptions directly from legal documents and classifying based on them,
leads to ambiguous results. For example, looking up the term Health at commercial classification systems such as Alexa~\cite{alexaTop} and SimilarWeb~\cite{similarweb}, we obtain domains about sports, healthy living, and healthy foods, as well as domains about chronic and sexually transmitted diseases. Manually distinguishing the truly sensitive among the not so sensitive ones, is not a scalable approach.




Even if one could compile large lists of truly sensitive domains, extracting and identifying the trackers operating on them remains a formidable task. This is due to problems such as distinguishing tracking from non-tracking third party domains~\cite{Roesner2012, openWPM-englehardt2016census, Nataliia2019}, complex recursive mechanisms used by trackers to invoke one another along multi-hop tracking chains, or active efforts by many of them to avoid programmatic detection and blocking~\cite{Mathur2018}. 


%
%
%
%
%

\subsection{Contributions}


We develop a scalable and accurate methodology for classifying sensitive domains across different categories,
including those mentioned in GDPR. Our approach requires a small manual effort to pick categories that are truly
sensitive from Curlie.org~\cite{curlie}, a large collection of URLs with annotated categories provided by a global
community of volunteer editors. This can be done easily due to the hierarchical nature of Curlie.org. Scanning, 
for example, the Health branch for truly sensitive sub-branches takes us less than five minutes. Having done that, we
immediately have access to thousands of well-labeled domains that can be used as training sets for deriving different
Machine Learning (ML) classifiers. The resulting classifiers can then be used to detect sensitive domains among
arbitrary sets of unlabeled domains, such as Top-K lists according to popularity. Having such classifiers is, thus, an
important building block for tracking services that track people on sensitive domains on the
Web.
for the benefit of other researchers and 


Moving on, we draw upon recent state-of-the-art work~\cite{Leung_2016, Mobile-Apps-NDSS2018, Binns2018} to implement a
powerful methodology for detecting third party trackers operating on large collections of sensitive domains. Our
methodology is able to detect complex, multi-layer inclusion of trackers~\cite{Arshad2017}, as well as the methods used
for exchanging information among them. We apply our detection methodology upon the sensitive domains that we used from Curlie.org to train our classifiers.


\subsection{Findings}

Analyzing around 30k domains that belong in sensitive categories, almost exclusive in the five sensitive categories as defined by
GDPR, we conclude that: 

\begin{itemize}

\item Our carefully filtered sets of sensitive domains are indeed tracked by a multitude of entities. The median number
of third party trackers found on sites of sensitive categories such as Health, Political Beliefs, and Sexual Orientation, is 10,
7, and 6, respectively. This median is smaller than the corresponding one among TopK domains (17), yet alarmingly high given the nature of these sites. 

\item On the top positions in terms of coverage we find the same trackers present at non-sensitive domains, which is again surprising given the very sensitive character of the domains we analyse.

\item We examine in more detail how popular trackers get into those domains, \ie whether they are intentionally present by being included  directly by first party domains, or appear unintentionally through recursive inclusions initiated by other third parties. We find that in the majority of cases the presence is intentional. 


\item Going beyond the mainstream trackers, we identify several niche trackers that are absent from non sensitive
domains but have clear presence on sensitive ones. Investigating them further, we find several ones that focus and
advertise on their web-sites their ability to track particularly sensitive categories. 

\item We study the communication patterns through so-called cookie
synchronization~\cite{Bashir2016,cookies-sync2019,castelluccia_2014} between mainstream and niche trackers and find that
the two often exchange information. We discuss potential consequences of allowing niche trackers access information held
by mainstream ones.   

 
\end{itemize}

%% file: sections/methodology.tex
\begin{figure*}[!bpt]
\includegraphics[width=.92\textwidth]{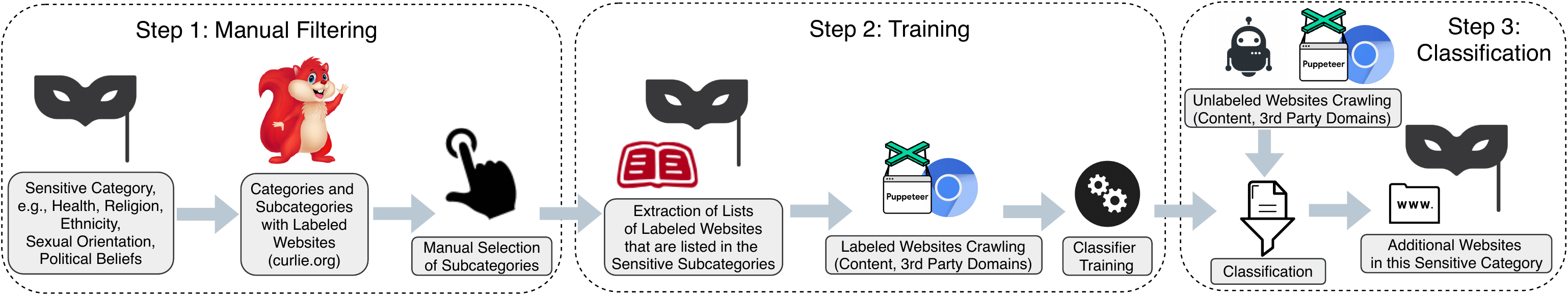}
\caption{Flowchart of our Methodology.}
\label{fig:methodology-flowchart}
\end{figure*}

\section{Classification Methodology}\label{sec:methodology}

In this section, we present, in detail, the methodology we develop to annotate websites that belong to sensitive
categories. Data protection laws define sensitive categories in generic high level terms such as, Health, Ethnicity, Religion,
Sexual Orientation, Political Beliefs (in the case of GDPR), Adult,
etc. Our methodology allows automating an additional refinement step that takes us from generic definitions of sensitive
terms to deciding whether a certain domain in the above categories is indeed sensitive, in the sense that most people
would rather not be seen visiting it. Our methodology requires a minimal manual fine tuning (something in the order of 5
minutes) and after that it is capable to accurately train machine learning classifiers that can be used to detect truly
sensitive domains in arbitrary lists of domains (\eg in TopK lists from Alexa~\cite{alexa-front} and other taxonomy systems). Combining this with our automated methodology for extracting tracking parties from lists of domains, opens up the road to putting in place a fully automated process for identifying sensitive domains and the trackers operating on them. In this paper we do this for analysis purposes but it is conceivable that our methods can also be used for proactive monitoring and enforcement.  

Figure~\ref{fig:methodology-flowchart} depicts a summary of our methodology. At a very high level, the methodology
involves three steps: Step 1 (Manual filtering): An expert user picks from a crowdsourced taxonomy of domains the
branches below a generic sensitive term such as Health that seem to be truly sensitive. Step 2 (Training): From the
selected branches we retrieve thousand of domains and use them to train a classifier. Step 3 (Classification): We apply
the trained classifier to arbitrary lists of domains in which we want to detect other sensitive domains beyond the ones
used in our training set. Next, we elaborate on the details of the above three steps. 


\subsection{Utilizing Labeled Websites}\label{sec:labeled-websites}

To achieve the desired refinement, we need a rich enough taxonomy and a fast way to pick truly
sensitive categories from it. We decided to use Curlie~\cite{curlie}, one of the largest human-edited directories of
the Web.  Curlie relies on \emph{category editors}, \ie experienced editors who specialize on a finite set of categories. This group of
editors makes up the majority of the Curlie community (around 92k active editors).
All \emph{new editors} apply to edit in small categories at first, and then apply to edit additional areas after they
have accumulated a number of edits. Community's senior editors are responsible for evaluating new editors' applications.
This ```Wikipedia'' style of indexing, helps in assuring a high quality labeling of URLs. 
Curlie contains more than 3.5 Millions annotated websites. These websites are full path URLs, not
just second level domains like in Alexa~\cite{alexaTop} and SimilarWeb\cite{similarweb}. This is a big advantage, as the
frontpage of a website is not always representative of the entire content of a large and complex domain. For example, the frontpage of a
University may not be annotated as a sensitive domain, but a website that is hosted in the medical school that describes
methods to treat cancer could be annotated as sensitive. Curlie, historically known as the Open Directory Project (ODP)
and DMOZ\cite{dmoz-odp} directory, is open source and free for use by all. Other directories are either not available to the public, \eg Google
AdWords~\cite{google_display_planner}, or are subscription-based and require payment, \eg Alexa and SimilarWeb. Although it is believed that
these commercial directories are also human-edited, there is no information on their annotation methodologies and way of work. Also,
several taxonomies related to commercial campaign planners avoid including very fine grained sensitive terms. For the above reasons we have picked Curlie. 

Typically, for each category, there is only one category that matches -- the first-level category. For each first-level category, there are a
few second-level categories and tens of third-level categories for each second-level category.  For example,
for the category Health, there is one first-level category, also called Health, and 47 second-level categories.
The user of our methodology, has to manually
select the first-level category, and then for the selected one, the second-level categories that are considered
sensitive. This is the only
manual component in our methodology, and the selection process takes less than 5 minutes for each (generic) sensitive
category. Studies have shown that involving humans at this stage can significantly improve the quality of the follow-up
automated process~\cite{The-Wisdom-of-the-Few}.  For example, in the case of Health, questionably sensitive
second-level categories such as Education, Employment, Animal, Healthcare Industry, are not selected, whereas
second-level categories such
as Conditions and Diseases, Addictions, and Mental Health are selected. Following this,  a list of labeled sensitive
websites becomes readily available by selecting all the websites under the kept second-level categories (all levels).
For example, in the case of Health, we manually selected six second-level sub-categories and this gave us 7,144 unique
URLs from 3,989 domains of truly sensitive Health related pages. The exact list will be made available upon publication together with all the other datasets and classifiers developed for our work. Even without releasing the data, one can browse the online version of Curlie to verify that many sub-categories of Health are full of truly sensitive domains.

\subsection{Crawling}\label{sec:crawling}

We have built a fully automated crawler using Google Puppeteer \cite{google_puppeteer}, 
NodeJS~\cite{nodejs}, and Google Chrome~\cite{chrome_browser}, and used it to visit and render all the websites that have been
selected in the previous step.  If a website is not available after three attempts or the crawler receives an
\textsf{``HTTP 404 - Page not Found''} error for a specific URL, then the website/URL is discarded. 

For each visited website the crawler collects two types of information. 
First, the crawler collect the full HTML code of the website, and second, all the HTTP(S) requests that are triggered during rendering time.  
Towards that end, the crawler utilizes a website scrolling functionality in order to trigger additional first and third party requests that are only initiated when a specific portion of the website is within the visible area of the web browser. 
This can happen due to lazy-loading~\cite{ChenZJH15} optimization algorithms. 
In the same path, we also impose an additional delay (1 minute) after the \textsf{window.onload()} event is triggered in order to give time to any tracking scripts and advertisements (if any) to fully load before we collect data. Note that since we operate at the browser level we can also observe and record encrypted (HTTPS) requests.

In the following sections we will explain how we use the collected data (a) to train a classifier to identify websites belonging to sensitive categories, 
and (b) to quantify the presence of third party domains in websites belonging to such categories.  

\subsection{Selecting and Training a Classifier}\label{sec:classifier}

Our goal in this paper is not to propose a new classification algorithm or suggest improvements to existing
classification methods, but rather to combine well known and well understood classification algorithms and off-the-shelf
tools~\cite{Complexity-vs-Performance-ML-IMC2017} with the right set of data to produce an automated methodology for identifying sensitive domains across different
categories. 

A number of existing classification algorithms are suitable for text classification. Some examples are, k-nearest-neighbor (KNN)~\cite{Han2001, Kwon2003, Calado2003}, Na\"ive Bayes~\cite{Domingos1997, Krishnaveni2016, Denoyer2004}, support vector machines (SVM)~\cite{Chakrabarti2003, Chen2006, Sun2002, Zhang2004}, decision tree (DT)~\cite{Wang2018, Estruch2006, YongHong2003}, neural network (NN)~\cite{Manevitz2007, Howard2010} and different variations~\cite{Meng2018}, maximum entropy~\cite{Chieu2002, Kwon2003}, \etc

We opt for using a Na\"ive Bayes classification algorithm. To be more specific, a multinomial Na\"ive Bayes algorithm, a
single supervised learning classifier that can predict multiple classes. Our choice relies on the following
observations. First, it has simple and easy training and classification stages~\cite{Sun2002}. Second, it is a fast
learning algorithm that can handle large numbers of features and classes~\cite{Mahesh2015, Chakrabarti2003}. Third, the algorithm has
already been tested and proved to work well using the old version of Curlie (DMOZ) categorization database
\cite{Oppenheimer2015}. Fourth, it has shown comparable,  and in some cases, better performance compared to other simple and
easy to use classification algorithms~\cite{Ting2011, Adetunji2018}. Finally, multiple off-the-shelf
implementations are publicly available for different programming languages and frameworks, thereby easily allowing other researchers to reproduce and validate our
results. 

\subsubsection{Input selection and pre-processing}\label{sec:classifier_input}
In Section~\ref{sec:labeled-websites} we explained how we collected our labeled dataset. 
In this section we provide more details on how we use it as a training input to our classifier.

For each visited website we have two source of information, the full HTML code and the list of HTTP(S) requests as observed during the rendering time of the page. 
From the HTML code we exclude all the non-visible elements (JavaScript, CSS, etc.) except the HTML \textsc{<META>} tag. From all the visible elements (human readable content) we extract all the text. 
We call this input source, the website \textit{content} (C). 
The \textsc{<META>} tag contains information related to the website, such as, the title of the page, a short description and keywords describing the page content. 
We call this input source \emph{meta-data} (M).

Using the HTTP(S) requests, we also consider a third source of input, that is, all HTTP(S) requests towards domains other than
the one that our crawler is actually visiting (a.k.a, third party domains). We call this input source \emph{third-party
domains} (TPD).  We also consider the third party domains that are present via recursive inclusion.  In this case, we
use the domain of the third party as well as the level of recursive inclusion as part of the domain name (\ie
domainA.com-0, domainB.com-1, \etc). We call this input source TPD-LVL. Later, we will examine if classification based on
meta-data or presence of third-party domains can be as precise as based on content. 

In the case of the two input sources, content and meta-data, we apply some standard preprocessing steps, such as, 
(1) language detection in order to exclude any websites with non English text, 
(2) transform all letters in lower case format and (3) remove any stop words (\ie a, the, \etc). 
Finally, (4) we set a minimum word length to three letters, numbers, or any combination of the two. 
In the case of the TPD and TPD-LVL input no preprocessing is required since we only have a list of domain names and not a free form text.

\subsubsection{Feature engineering and training}\label{sec:feature_eng}
With respect to feature selection, we consider two algorithms as follow:

\noindent{\bf Bag-of-Words} (BoW): It is a popular Information Retrieval (IR) technique to represent text as a multiset of its
words by disregarding grammar, but, keeping multiplicity, \ie the number of appearances of a word in a document or a corpus of documents.

\noindent{\bf Term Frequency-Inverse Document Frequency} (TF-IDF): It is a popular IR numerical statistic that is intended to
reflect how important a word is to a document. The TF-IDF value increases proportionally to the number of times a word
appears within a document and inversely proportionally to its corresponding frequency among other documents.

During this step each feature selection algorithm searches through all possible combinations of attributes in the data
(word term in our input) to find which subset of features (words) works best for prediction. We use 70\% of each dataset
during the training phase and the remaining 30\% during the validation phase.

\subsection{Classifying Unlabeled Websites}\label{sec:unlabeled}

We can use the trained classifier to identify additional websites belonging to the number of classes that our classifier
is trained to identify.  In order to achieve that, we can use any list of websites that we want to examine.  Then, we need to
follow the methodology described above, that is, use a crawler to fetch the HTML code of the websites and apply the
preprocessing step.  Depending on the number of classes that our classifier is able to identify, during the prediction
phase the classifier returns a vector of probabilities between 0 and 1 with equal length as the number of classes that
it can identify. The position with the highest probability \textit{``P''} defines the prediction category with the
corresponding probability score \textit{``P''}. If the website being classified does not belong to any of the sensitive categories that the classifier is trained to find, the probability vector will include only small values. 


%% file: sections/classification.tex
\section{Classifier Evaluation}\label{sec:results}

In this section, we assess the accuracy of classifiers trained by labeled data as described in Section~\ref{sec:classifier}.
We first describe in more detail the different datasets we used. Then, we evaluate the accuracy of different classifiers
on those labeled data as well as their sensitivity to different parameters. Lastly, we assess the accuracy and robustness
of the classifiers when applied to unlabeled domains from TopK lists.

\subsection{Datasets}\label{sec:datasets}

\begin{table}[]
\caption{Statistics about the sensitive subcategories and labeled websites we used in our study.}
\label{tab:labeled-per-category}
\resizebox{\columnwidth}{!}{%
\begin{tabular}{|c|c|c|c|c|c|}
\hline
Category   & \# Selected 1st-& \# Labeled Websites & \# Labeled & \# Domains &\# Labeled\\
           & (2nd-) level & in Selected 2nd-     & Websites after & &Websites after\\
           & Categories & level Categories    & Crawling & & Filtering\\ \hline \hline
Health         & 1 (6)   & 7,164  & 7,144  & 3,989 & 4,909 \\ \hline
Ethnicity      & 1 (21)  & 3,127  & 3,126  & 2,299 & 1,985 \\ \hline
Religion       & 1 (42)  & 10,145 & 10,134 & 7,018 & 5,726 \\ \hline
Sexual Orien.  & 1 (10)  & 3,373  & 3,373  & 2,787 & 1,956 \\ \hline
Political Bel. & 1 (15)  & 5,547  & 5,516  & 3,874 & 3,201 \\ \hline
Porn           & -       & 1,301  & 1,301  & 1,293 &   781 \\ \hline
\end{tabular}
}
\end{table}

To bootstrap our study we consider the five generic sensitive categories according to GDPR~\cite{EU-GDPR}, namely,
Health, Ethnicity, Religion, Sexual Orientation, and Political Beliefs (the category Biometric Data is not applicable to
our study). We identify the relevant categories in Curlie and by allocating less than five
minutes for each category we select the first-level and second-level categories. Then, we consider all the labeled websites for each
of the five categories. For the statistics of the number of first-/second-level categories and labeled websites per category, please see
Table~\ref{tab:labeled-per-category} (second and third column). 

Following the methodology described in Section~\ref{sec:classifier_input}, first the orchestrated crawler downloads all
the websites from each category, and second, we pre-process the downloaded data. The result of this process is
summarized in Table~\ref{tab:labeled-per-category} (fourth column) for each sensitive category. Note that Curlie provides
the full URL, for the number of domains per category, see Table~\ref{tab:labeled-per-category} (fifth column). The
crawling took place between March and April 2019. For each website, the corpus contains the
human-readable text, the meta-data, and all the third party domains as we have already explained in
Section~\ref{sec:classifier_input}.  

\eat{
Then, our orchestrated crawler downloads all the websites per category, and our data pre-processor filters the content
of the webpage as described in Section~\ref{sec:classifier_input}. After this filtering, our corpus is as summarized per category
in Table~\ref{tab:labeled-per-category} (fourth column).  
}

{\bf Ethical considerations:} Due to the sensitive content of the websites, we decided not to use any real user or any
personal identifier. The crawlers utilized IPs assigned to cloud providers and universities. Note also that since we crawl using scripted web-browsers we see all the content and HTTP(S) headers and don't have any issues due to transport layer encryption. 

\subsection{Classification Accuracy}\label{classification-accuracy}


{\em Classification accuracy} is defined as the percentage score, from 0\% (lowest) to 100\% (the highest), that a
classifier can accurately assign websites to the associated category. In this section, we investigate how different
inputs (features), combinations and parameters can influence our classifier's accuracy.


As a first step, we have to assess which feature sets (available input options) are suitable for our classification task. Towards that end, and to avoid any bias, we consider an equal number of websites for each category. 
We also set the number of features to be three thousand in order to evaluate the different input options (we vary this later). The category
with the minimum number of websites is Sexual Orientation with 1,956 (for the other categories, see
Table~\ref{tab:labeled-per-category} (last column)). Thus, for each of the other categories, we uniformly at random select 1,956 websites.

\eat{
\begin{table}[t]
\caption{Overview of the accuracy for different feature set sources and feature types (Bag-of-Words and TF-IDF). 
Classifier parameters: English language, lowercase text, 3k features, minimum three characters per word.}
\label{table:overview-classifier-results}
\resizebox{\columnwidth}{!}{%
\begin{tabular}{|c|c|c|}
\hline
Feature Source & BoW & TF-IDF\\ \hline \hline
 Content (C) & 79.92\% & 84.29\%\\ \hline
 Meta-data (M) & 77.92\% & 77.10\% \\ \hline
Third-party domains (TPD) & 43.28\% & 44.26\%\\ \hline
 TPD with Levels (TPD-LVL) & {\color{red} 42.27\%} & 43.19\% \\ \hline
 M + C & 81.50\% & {\bf \color{blue} \textbf{85.65\%}}\\ \hline
 M + C + TPD & 81.44\% & 85.59\%\\ \hline
 M + C + TPD-LVL & 81.47\% & 85.62\% \\ \hline
\end{tabular}
}
\todo{add Feature Engineering above BoW and TF-IDF columns}
\end{table}
}

\begin{table}[]
\caption{Overview of the accuracy for different feature set sources and feature engineering options. Classifier parameters: English text, lowercase, stop words removal, 3k features, minimum three characters per word.}
\label{table:overview-classifier-results}
\begin{tabular}{c|c|c|}
\cline{2-3}
                                                & \multicolumn{2}{c|}{Feature Engineering} \\ \hline
\multicolumn{1}{|c|}{Feature Source}            & BoW                 & TF-IDF             \\ \hline \hline
\multicolumn{1}{|c|}{Content (C)}               & 79.92\%             & 84.29\%            \\ \hline
\multicolumn{1}{|c|}{Meta-data (M)}             & 77.92\%             & 77.10\%            \\ \hline
\multicolumn{1}{|c|}{Third Party domains (TPD)} & 43.28\%             & 44.26\%            \\ \hline
\multicolumn{1}{|c|}{TPD with Levels (TPD-LVL)} & {\color{red} 42.27\%}             & 43.19\%            \\ \hline
\multicolumn{1}{|c|}{M + C}                     & 81.50\%             & {\bf \color{blue} \textbf{85.66\%}}            \\ \hline
\multicolumn{1}{|c|}{M + C + TPD}               & 81.44\%             & 85.59\%            \\ \hline
\multicolumn{1}{|c|}{M + C + TPD-LVL}           & 81.45\%             & 85.62\%            \\ \hline
\end{tabular}
\end{table}

Following our results depicted in Table~\ref{table:overview-classifier-results}, when the classifier extracts features only from the text content of the webpage (C), the classification accuracy is
quite high. The accuracy with BoW is close to 80\% and with TF-IDF above 84\%. Using only the website meta-data (M) as feature source 
yields high classification accuracy, but significantly lower than when considering the text of the website both with BoW
and TF-IDF. Third party domains, used directly (TPD) or annotated with the inclusion level (TPD-LVL) yield very bad
classification accuracy, for both BoW and TF-IDF -- close to half of the accuracy when considering the text content of the
webpage (C). We also tested all the different
combinations of feature sources, and as shown in Table~\ref{table:overview-classifier-results}, the combination of
website text content and meta-data with TF-IDF yields the best classification accuracy (85.66\%). 

\begin{figure}[t]
	\centering
\subfloat[Content and Meta-data with BoW.]{{\includegraphics[width=0.49\columnwidth]{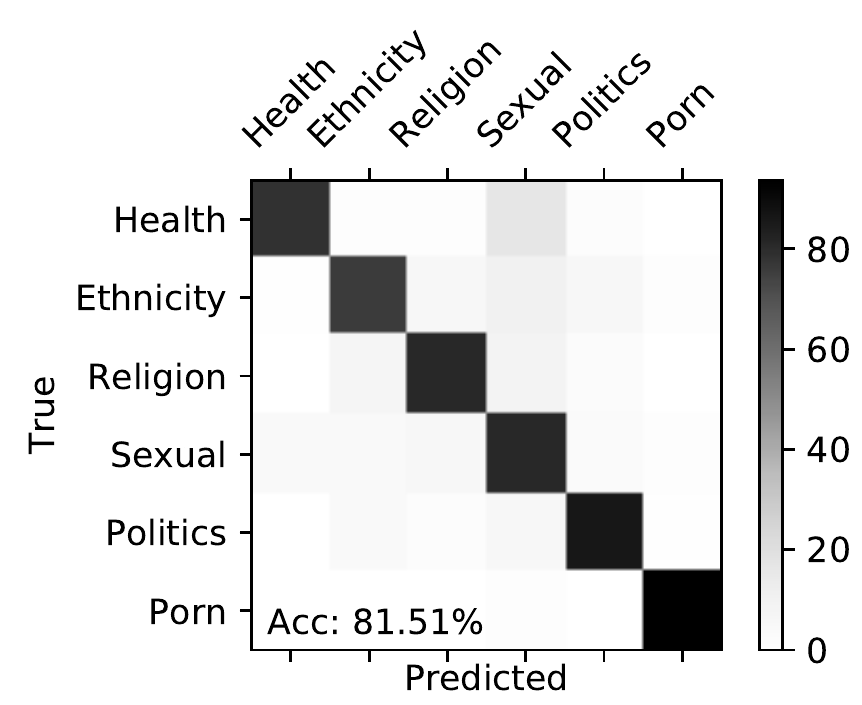}}}\hfill
\subfloat[Content and Meta-data with TF-IDF.]{{\includegraphics[width=0.49\columnwidth]{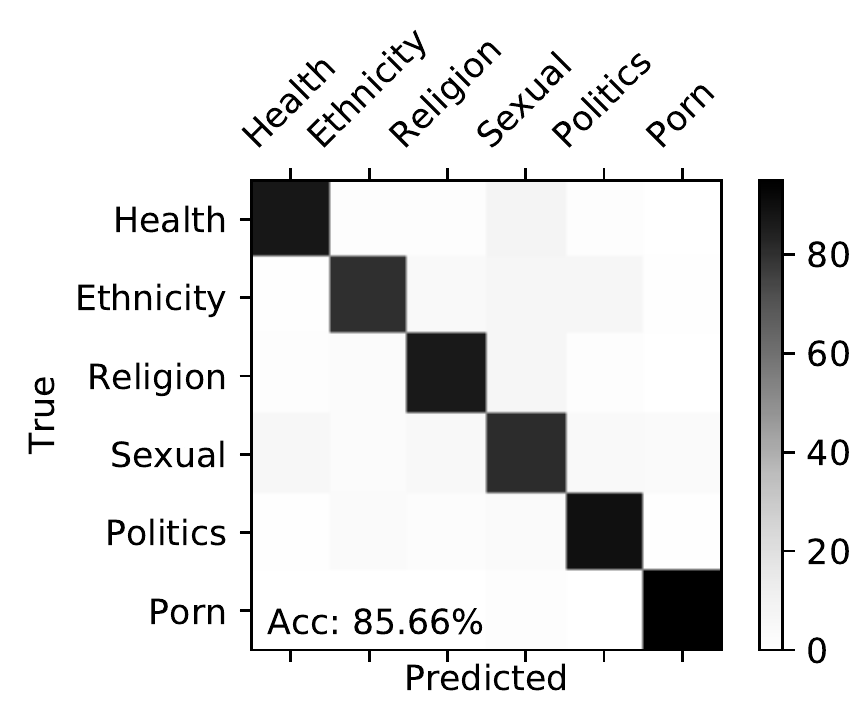} }} 
\caption{Confusion matrix for the sensitive categories using as feature source the website text content and the meta-data as as feature types the BoW and the TF-IDF.}
\label{fig:confusion-matrix-content}
\end{figure}

\subsubsection{Accuracy per category}\label{sec:accuracy-category}

We then turn our attention to the classification accuracy per category. Notice that we use only a subset of the webpages
in the classifier, thus, we would like to assess if the classifier is able to correctly classify the websites to the
appropriate category. In Figure~\ref{fig:confusion-matrix-content} we present the confusion matrix, i.e., table layout
that allows visualization of the performance of our classification.  Each raw of the confusion matrix represents the
percentage of the instances in the actual class, while each column represents the percentage of the instances in a
predicted class (sensitive category).  A high-level observation is that the classifier predicts with high accuracy the
class of the websites that belong to the same category (the darker the cell the better).

Nevertheless, there are some shadows that indicate that in some cases the classifier mis-predicts the class. 
After closely investigating the different categories, we observe the following. 
For example, in the sexual orientation category, there are
multiple websites (and Curlie subcategories) that are associated
to a specific type of sexuality, e.g., websites dedicated to only gay men or lesbian women,  that do not share a lot of similar words in the webpage text (i.e. gay website use men related words and lesbian websites use women related words). In the case of Health and Sexual Orientation, we observe that in some URLs related to gay also discuss sexually transmitted diseases. Overall the category where the classifier achieves the highest accuracy is Political Beliefs (88.3\%) and the category with the lowest accuracy is Sexual Orientation (74.2\%). 

\subsubsection{Feature sets}

\begin{table}[t]
\caption{Top 10 features (out of the 3k) per sensitive categories. 
Content and meta-data are the feature sources and the TF-IDF algorithm is used for feature engineering (only English
language websites).}
\label{table:top10-features}
\resizebox{\columnwidth}{!}{%
\begin{tabular}{|c|c|c|c|c|c|c|}
\hline
Ranking & Health & Ethnicity & Religion   & Sexual Or. & Politics   & Porn  \\ \hline\hline
1   & Cancer     & Native    & Church     & Gay        & Party      & Porn  \\ \hline
2   & Disease    & Indian    & God        & Lesbian    & Democrats  & Sex   \\ \hline
3   & Health     & American  & Bible      & Sex        & News       & Video \\ \hline
4   & Treatment  & Language  & Christian  & LGBT       & Government & Teen  \\ \hline
5   & Syndrome   & Tribe     & Faith     & Queer      & State      & Free  \\ \hline
6   & Symptoms   & History   & Prayer     & Bisexual   & Political  & Girls \\ \hline
7   & Patients   & People    & Holy       & Community  & Liberal    & Anal  \\ \hline
8   & Research   & Culture   & Catholic   & HIV        & Election   & Pussy \\ \hline
9   & Clinical   & Tribal     & Religious & Pride      & Parliament & Hot   \\ \hline
10  & Care       & Indigenous & Love      & Events     & Labour     & Big   \\ \hline
\end{tabular}
}
\end{table}

Another indication that the classification is robust is the obvious relevancy of the features with the highest weight selected by the classifier. In
Table~\ref{table:top10-features} we list the top 10 features (out of the 3k) in each sensitive category.
The classifier is using as an input the websites text content and meta-data as well as the TF-IDF algorithm for feature engineering.
We observe that the keywords (top features), that are automatically generated by the classifier, are well suited to
characterize each one of the sensitive categories.  

\subsubsection{Extending the Classifier beyond GDPR sensitive categories}\label{sec:extended-classifier}

Another advantage of the classifier is that as new categories are added as an input, it is possible to provide
additional training and improve it. For this, we added as input another sensitive category, namely, Porn. All the
quality characteristics of the classifier, such as, classification accuracy, list of features, and the confusion matrix were slightly
improved. We build the list of Porn websites by manually selecting them from different blocking
lists~\cite{filterlists} related to adult content websites.

\subsection{Sensitivity on the number of features}\label{sec:number-features}

\begin{figure}[t]
\centering
\subfloat[Content and Meta-data with BoW.]{{\includegraphics[width=\columnwidth]{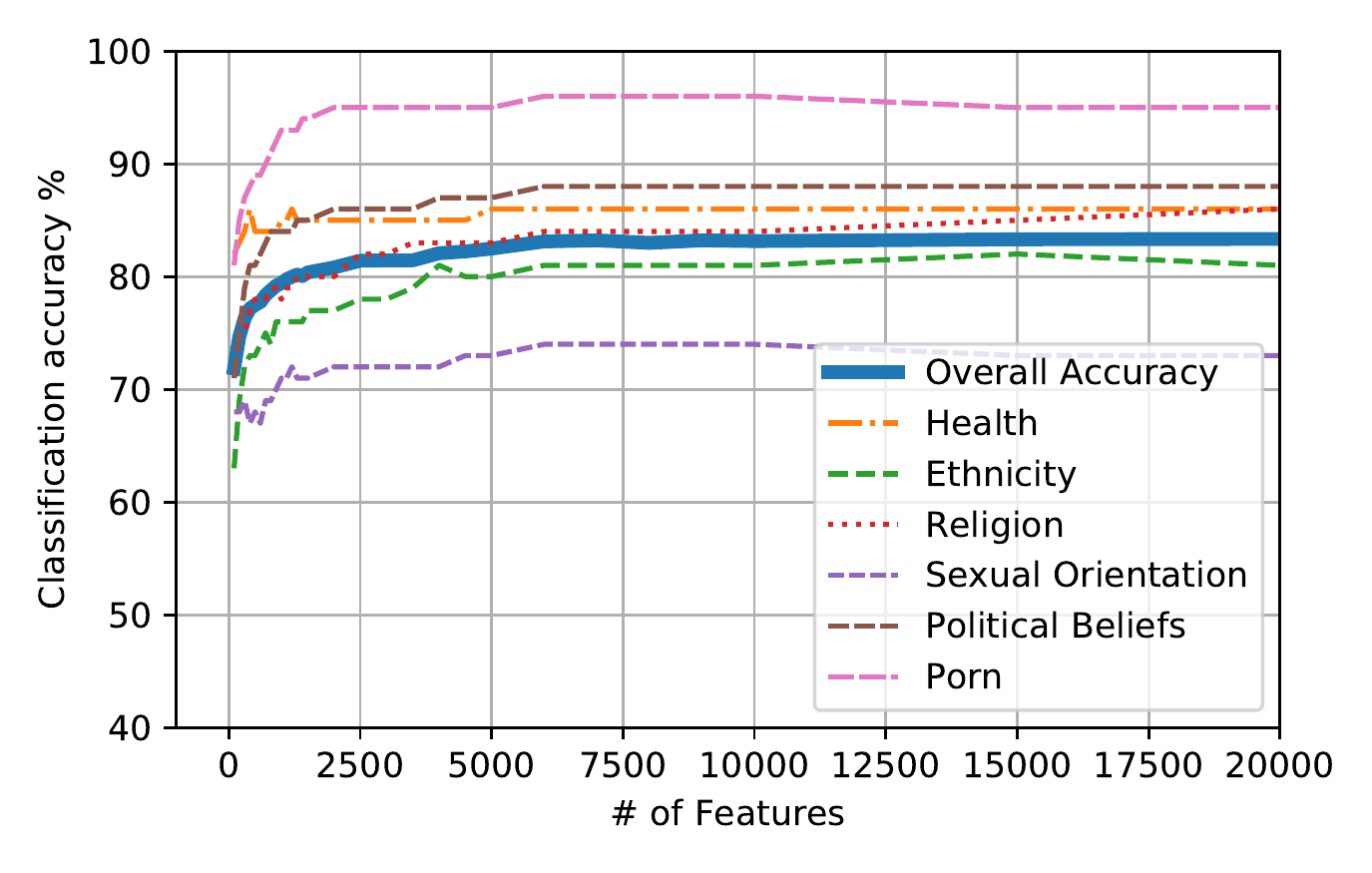}}}\hfill
\subfloat[Content and Meta-data with TF-IDF.]{{\includegraphics[width=\columnwidth]{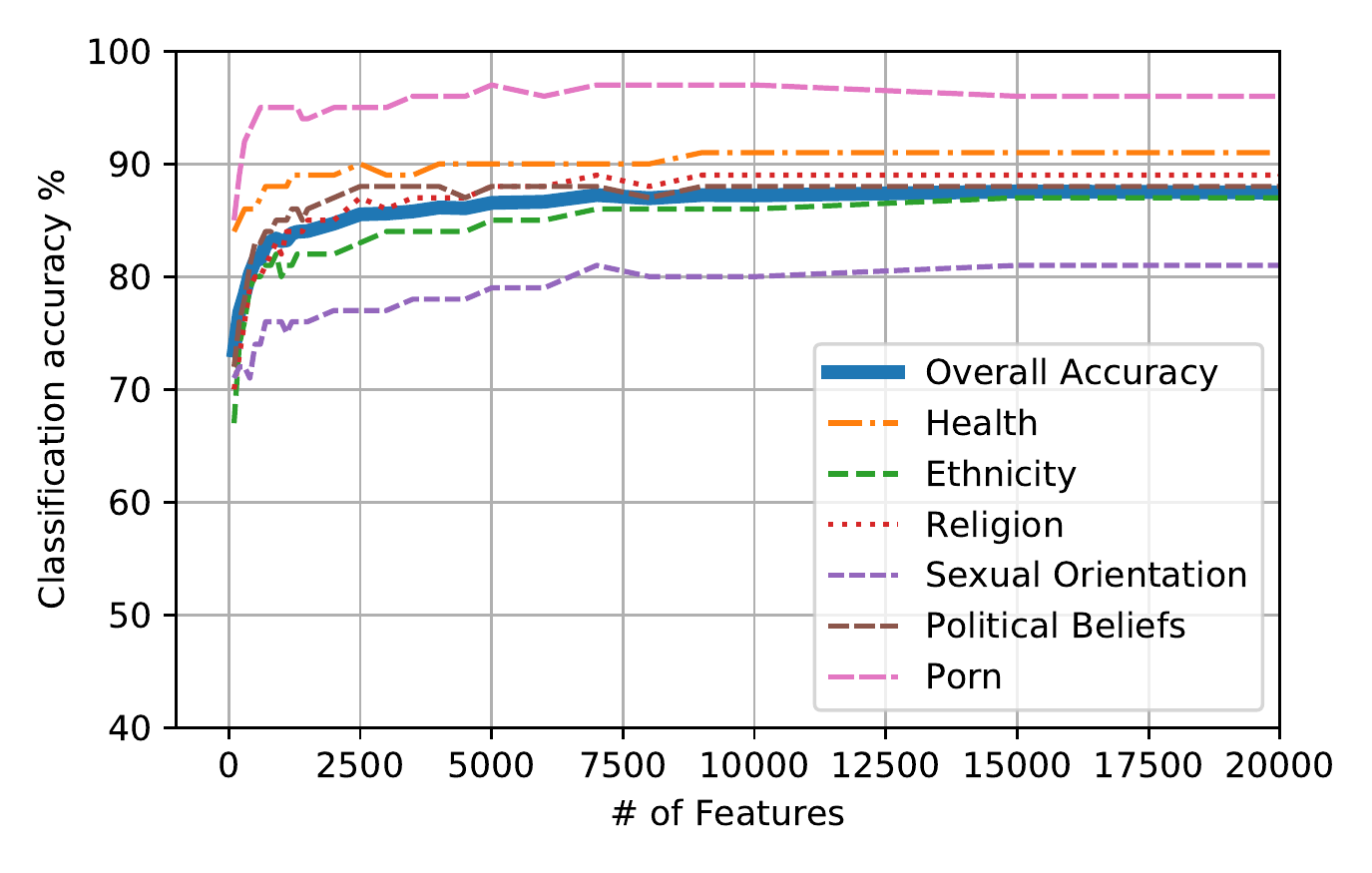}}}
\caption{Classification accuracy as the number of features increase with BoW and TF-IDF.}
\label{fig:effect-features}
\end{figure}

Next, we examine the effect of the number of features used in the training of the classifier on the classification accuracy.
Figure~\ref{fig:effect-features} shows the classification accuracy for each sensitive
category as well as the overall accuracy, as the number of features increases. 
We present the results for Bag-of-Word and TF-IDF as feature engineering algorithms, see Figure~\ref{fig:effect-features} (a) and (b), respectively.
For each sensitive category and feature engineering algorithms, the classification accuracy increases with the number of features. 
Our analysis shows that, for each and all categories, there is a ``knee'' in the classification accuracy when the number of
features is around 1,000. From 3,000 features and onwards, the classification accuracy improvement is marginal. One important observation is
that the converged classification accuracy varies across sensitive categories, but consistently, TF-IDF outperforms BoW
across all categories. For the rest of the results presented in this paper, unless otherwise noted, we will use 3,000
features with textual content and meta-data information, and with TF-IDF for our classification.

\begin{table}[!bpt]
\caption{Execution time for feature engineering and training time using different combination of inputs and number of
features.}
\label{table:execution_times}
\resizebox{\columnwidth}{!}{%
\begin{tabular}{|c|r|c|c|c|c|}
\hline
\multirow{2}{*}{Input}       & \multicolumn{1}{c|}{\multirow{2}{*}{\# Features}} & \multicolumn{2}{c|}{Feature Eng. Time} & \multicolumn{2}{c|}{Training Time} \\ \cline{3-6} 
                             & \multicolumn{1}{c|}{}                             & BoW               & TF-IDF             & BoW              & TF-IDF          \\ \hline
\multirow{3}{*}{Content (C)} & 1,000                                             & 13.4s             & 12.9s              & 28.1ms           & 21.7ms          \\ \cline{2-6} 
                             & 10,000                                            & 13.1s             & 12.9s              & 48.3ms           & 39.4ms          \\ \cline{2-6} 
                             & 100,000                                           & 13.2s             & 13.1s              & 91.3ms           & 79.4ms          \\ \hline
\multirow{3}{*}{Meta-data (M)}    & 1,000                                             & 920ms             & 359ms              & 18.6ms           & 16.5ms          \\ \cline{2-6} 
                             & 10,000                                            & 893ms             & 367ms              & 19.2ms           & 17.9ms          \\ \cline{2-6} 
                             & 100,000                                           & 878ms             & 368ms              & 26.4ms           & 23.8ms          \\ \hline
\multirow{3}{*}{M + C}       & 1,000                                             & 13.2s             & 13.0s              & 26.7ms           & 22.8ms          \\ \cline{2-6} 
                             & 10,000                                            & 13.2s             & 13.1s              & 44.4ms           & 32.8ms          \\ \cline{2-6} 
                             & 100,000                                           & 13.4s             & 13.4s              & 86.2ms           & 67.6ms          \\ \hline
\end{tabular}
}
\end{table}

With regards to the feature engineering and training time for the classifier, in Table~\ref{table:execution_times},
we report the execution time for the two 
best feature sources, namely, content and meta-data, as well as the execution time when both sources are used.
All the experiments were executed on a MacBook Pro 2018 version.
A first observation is that the feature engineering time for content and meta-data does not depend on the number of
features (ranging from 1k to 100k features) nor the feature engineering method. This is to be expected as the input is the same, i.e., the corpus of textual
content and meta-data. The feature engineering algorithm, however, does depend on the input source. Utilizing only meta-data
takes less than a second, where textual content takes about 13 seconds.
The training time does depend on the number of features and the feature engineering method, as shown in
Table~\ref{table:execution_times} (last two columns), but this time is almost negligible. 
For either feature engineering algorithm, the training time increases only three times, when the set of features increases by three orders of magnitude. 
Thus, training of the classifier is very scalable.
When we consider different feature engineering algorithms,  
TF-IDF not only yields better classification accuracy than BoW, but is also faster. 
The training of our
classifier with 3k features and TD-IDF takes only 23msec.

\subsection{Performance of the Classifier on unlabeled data}\label{sec:classifier-on-unlabeled-data}


Next, we evaluate the ability of the classifier trained with labeled sensitive domains from curlie.org to accurately
classify unlabeled sensitive domains in TopK popular lists. As a showcase, we use the Alexa Top 20k list.
We use the latest list on the day of crawling (March to April 2019).
We decide to focus only on the top 20k as many of popular webpages are there and the churn is relative
small~\cite{Alexa-stability:IMC2018}. After the crawling and applying filters for English
language and sufficient text and meta-information we maintained 7,115 websites.  We
remind the reader that the main motivation for the training of such classifiers is to use them to automatically identify
sensitive domains by filtering TopK lists and then use the identified domains to extract the trackers present on them
for analysis or for compliance audits.  

After running our classifier on the above list we identify several sensitive domains across the different categories.  We
limit the number of selected websites to 1k out of the 7,115 from Alexa in order to be able to manually examine if
indeed the predicted categories are correct or not.  When the prediction probability threshold is set to a small value,
such as 0.5 (50\%), the classifier annotates 186 as sensitive. In more detail, we get 44 websites belonging to the category
Porn, 15 related with Health, 64 with Political Beliefs, 9 with Sexual Orientation and finally 2 with Ethnicity. We manually
investigated and confirmed that 134 out of the 186 annotated by the classifier were indeed sensitive and belonging to
the corresponding discovered categories. 

Our next objective is to optimize the threshold in order to maximize the number of discovered sensitive websites, but
without mis-classifying many websites as sensitive (false positives). We remind the reader that our use case is to
collect lots of sensitive websites from arbitrary lists to extract who is tracking them. Thus,
we are more sensitive to the precision (positives being real) rather than the recall (finding all positives) of the
classification. 

\begin{figure}[t]
\center  
\includegraphics[width=\linewidth]{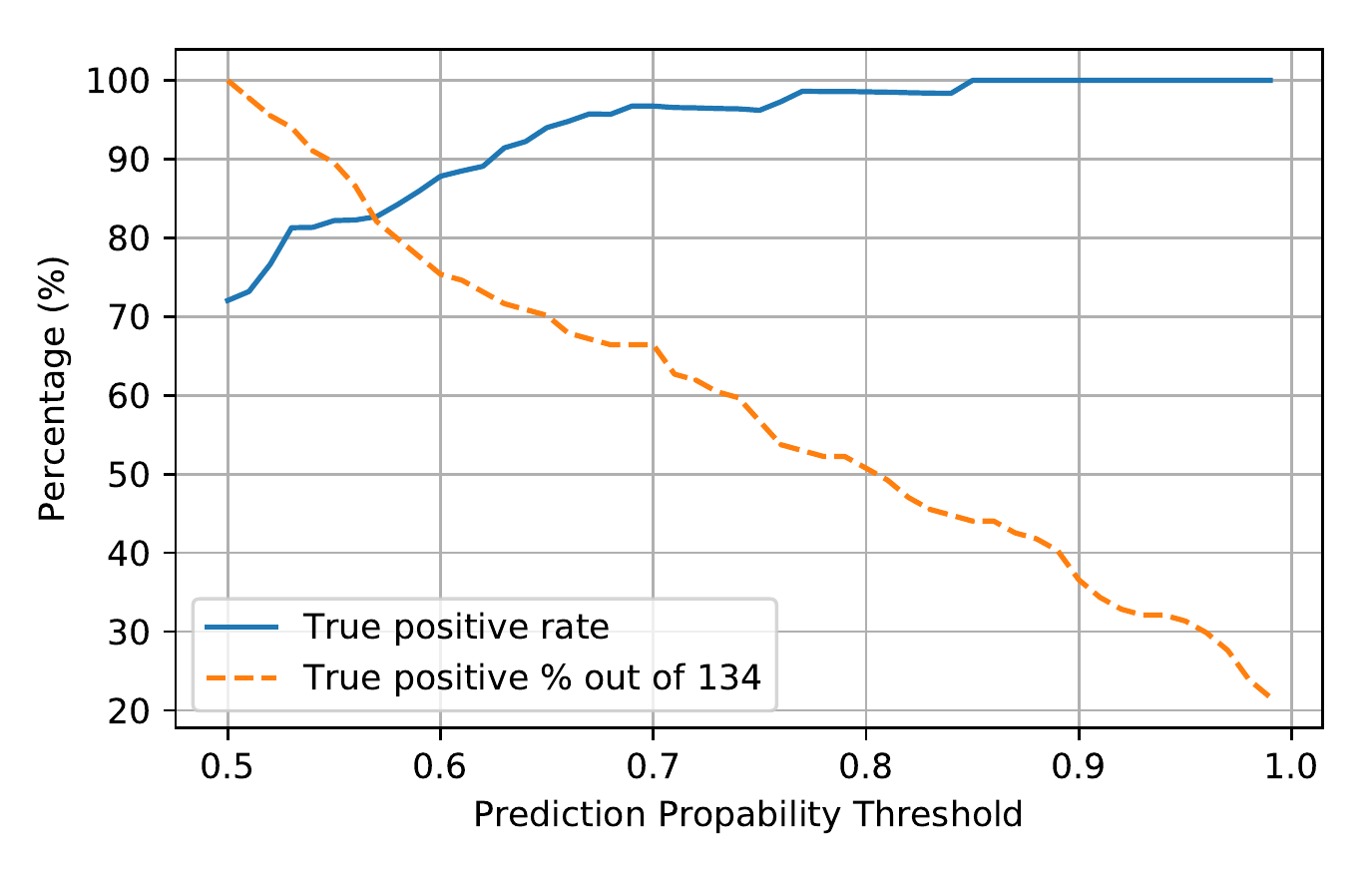}
\caption{The true positive rate and the percentage of a total (134) true positive sensitive websites out of a total 1k unlabeled website that we
examine as a function of the classifier's prediction probability threshold.}
\label{fig:classifierTopK_predictions}
\end{figure}

In Figure~\ref{fig:classifierTopK_predictions}, we show how the percentage of true positive rate increases as the
prediction probability threshold of the classifier increases. However, as the true positive rate increases, less
sensitive websites (as annotated by our manual investigation) are identified as such.
Moreover, the percentage of a total (134 out of a total 1k) true positive sensitive websites 
decreases slower than the true positive rate when probability threshold of the classifier increases. 
This suggests that we can achieve very high positive rate by missing a relative small number of true positive sensitive
websites. As shown in Figure~\ref{fig:classifierTopK_predictions}, the true positive rate exceeds 90\% (and more than
70\% of the sensitive sites are identified as such) when the
classifier's prediction probability threshold is 0.63. 

%% file: sections/trackers.tex
\begin{table*}[]
\caption{Statistics about the presence of third party domains per webpage at TopK and sensitive categories.}
\label{table:summary-3rd-parties}
\resizebox{\textwidth}{!}{%
\begin{tabular}{|l|r|r|r|r|r|r|r|r|r|r|r|r|r|}
\hline
\multicolumn{1}{|c|}{\multirow{3}{*}{Category}} & \multicolumn{1}{c|}{\multirow{3}{*}{\# Websites}} & \multicolumn{4}{c|}{\# 3rd party HTTP Requests}                                                                                  & \multicolumn{4}{c|}{\# Unique 3rd parties full domain}                                                                           & \multicolumn{4}{c|}{\# Unique 3rd parties TLD+1}                                                                                 \\ \cline{3-14} 
\multicolumn{1}{|c|}{}                          & \multicolumn{1}{c|}{}                             & \multicolumn{1}{c|}{\multirow{2}{*}{Total}} & \multicolumn{3}{c|}{Per Website}                                                   & \multicolumn{1}{c|}{\multirow{2}{*}{Total}} & \multicolumn{3}{c|}{Per Website}                                                   & \multicolumn{1}{c|}{\multirow{2}{*}{Total}} & \multicolumn{3}{c|}{Per Website}                                                   \\ \cline{4-6} \cline{8-10} \cline{12-14} 
\multicolumn{1}{|c|}{}                          & \multicolumn{1}{c|}{}                             & \multicolumn{1}{c|}{}                       & \multicolumn{1}{c|}{Median} & \multicolumn{1}{c|}{Mean} & \multicolumn{1}{c|}{STD} & \multicolumn{1}{c|}{}                       & \multicolumn{1}{c|}{Median} & \multicolumn{1}{c|}{Mean} & \multicolumn{1}{c|}{STD} & \multicolumn{1}{c|}{}                       & \multicolumn{1}{c|}{Median} & \multicolumn{1}{c|}{Mean} & \multicolumn{1}{c|}{STD} \\ \hline
TopK                                            & 7,115                                             & 898,929                                     & 79                          & 126.34                    & 152.23                   & 27,985                                      & 17                          & 24.78                     & 24.92                    & 11,524                                      & 12                          & 16.54                     & 15.92                    \\ \hline
Religion                                        & 10,134                                            & 497,284                                     & 18                          & 49.07                     & 77.31                    & 6,458                                       & 6                           & 11.78                     & 16.68                    & 3,619                                       & 5                           & 9.03                      & 12.76                    \\ \hline
Health                                          & 7,144                                             & 464,660                                     & 25                          & 65.04                     & 96.2                     & 5,821                                       & 10                          & 16.44                     & 18.92                    & 2,959                                       & 8                           & 11.66                     & 12.32                    \\ \hline
Political Beliefs                               & 5,516                                             & 411,279                                     & 32                          & 74.56                     & 115.60                   & 4,973                                       & 7                           & 14.65                     & 24.15                    & 2,559                                       & 5                           & 10.01                     & 15.51                    \\ \hline
Sexual Orientation                              & 3,373                                             & 156,658                                     & 19                          & 46.44                     & 97.53                    & 4,355                                       & 6                           & 10.28                     & 13.79                    & 2,424                                       & 5                           & 7.48                      & 9.12                     \\ \hline
Ethnicity                                       & 3,126                                             & 105,203                                     & 14                          & 33.65                     & 53.62                    & 3,327                                       & 5                           & 8.28                      & 10.80                    & 1,873                                       & 4                           & 6.14                      & 7.46                     \\ \hline
Porn                                            & 1,301                                             & 92,698                                      & 40                          & 71.25                     & 112.61                   & 3,060                                       & 7                           & 8.59                      & 8.0                      & 1,865                                       & 5                           & 6.07                      & 5.66                     \\ \hline
All Sensitive                                   & 30,594                                            & 1,727,782                                   & 22                          & 56.47                     & 92.73                    & 18,847                                      & 7                           & 12.73                     & 18.0                     & 10,140                                      & 5                           & 9.23                      & 12.34                    \\ \hline
\end{tabular}%
}
\end{table*}

\begin{figure}[!bpt]
\center  
\includegraphics[width=1\linewidth]{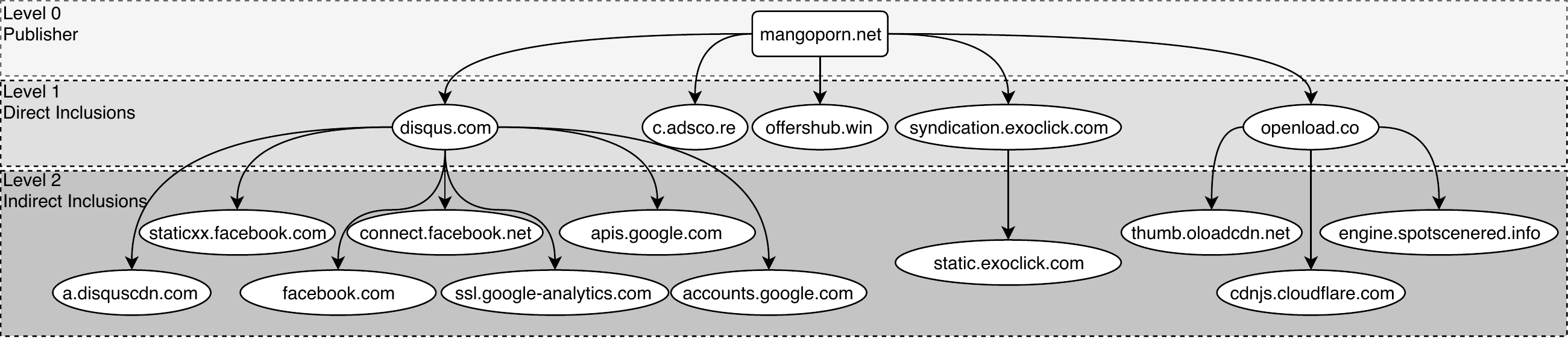}
\caption{Adult website rendering example and the inclusion chains of third parties at different inclusion levels.
}
\label{fig:chainsExample}
\end{figure}

\section{Trackers on Sensitive Domains}\label{sec:trackers}

In this section, we first present in detail our methodology for detecting the presence of trackers on particular
domains. Then, we quantify the amount of tracking that takes place on sensitive domains. Finally, we identify and
catalogize the tracking services operating on them. 

\subsection{Detection methodology}

{\bf Third party inclusion chains:}
For each one of the rendered pages belonging to sensitive domains crawled as described in the previous section, we conduct the following. 
First, we detect third party domains by following all requests towards any domain beyond the one originally visited by our crawler. In order to identify if a third party domain is a tracker or not, one can use different filter lists and methodologies as descriped in~\cite{Iordanou:GDPR-IMC2018, openWPM-englehardt2016census}. 
Nevertheless, we choose to include all third party domains in our dataset without any filtering in order to observe if any unknown third party tracker operate in sensitive category websites. As will be shown later, most of the third party domains encountered are indeed trackers.     

Next, using the collected data we construct the third-party request inclusions for each 
webpage visit. To do so, we combine information related to each individual third-party request, such as, the request 
type (\textit{i.e.,} the \textit{``sub\_frame''} type identifies an iFrame creations), the initiator URL (source) and 
request URL (destination) as reported by the \textit{``onBeforeSendHeaders''} event listener, the iFrame id and other meta-data information as necessary.  To correctly identify the referrer field of a request when a 
third-party script is directly embedded in the first-party domain content, we monitor all \textit{``sub\_frame''} 
requests executed at the first-party domain level~\cite{bashir2018}. We then assign the correctly inferred  
third-party domain to the newly created iFrame based on 
the request URL that is responsible for the iFrame creation.

An example of such inclusions is depicted in Figure~\ref{fig:chainsExample}. 
At level 0 (top box) is the actual
user-initiated visit to a specific first-party domain, in our example ``mangporn.net'' an adult content website. At
level 1 (middle box) there are 5 different third-party domains that are directly included by the first-party domain
(publisher) henceforth referred to as \textit{Direct Inclusions}. At level 2 (bottom box) there are 11 domains that are
included by 3 different third-party domains already included by the publisher at level 1.  We refer to such third-party
domain inclusions above level 2 as \textit{Indirect Inclusions}. Note that in our example, we use a website with only 2
levels of inclusion. In our dataset we have observed up to 9 levels.

To be able to understand the interaction between the different domains that we observe in each website, we introduce the
notion of \textit{Inclusion chains}. An inclusion chain is a path that connects domains based on the order of inclusion.
Using Figure~\ref{fig:chainsExample}, we can identify an inclusion chain between the first-party domain
``mangoporn.net'' towards the third-party domain
``a.discuscdn.com''. The chain includes the following domains in the exact order:
$\text{``mangoporn.net''} \rightarrow \text{``disqus.com''} \rightarrow \text{``a.disquiscdn.com''}$.
Overall in Figure~\ref{fig:chainsExample} we have 13 such inclusion chains. Note here that all these inclusions are not
visible in the static code of a page. They appear only when one fully renders the page.  For example, \text{nytimes.com}
and \text{nbcnews.com} render their content and ads dynamically while the user scrolls the page. We overcome this issue as we explain in section~\ref{sec:crawling}.


Table~\ref{table:summary-3rd-parties} summarized the number of sites per category, the total number of HTTP(S) requests in
our crawling, and the number of unique third parties.

\eat{
{\bf Detecting Trackers:} To identify Ad and Tracking related domains we use
AdBlockPlus~\cite{adblockplus} ``easylist'' and ``easyPrivacy'' lists~\cite{easylist} and a custom list of
keywords related to Ad and Tracking activities that we empirically build.
For more details related to the keyword list see Section~XXXX.
In total, we identify more than XXXK domain belonging to those two categories.
}

\begin{figure}[!bpt]
  	\centering{{\includegraphics[width=1\columnwidth]{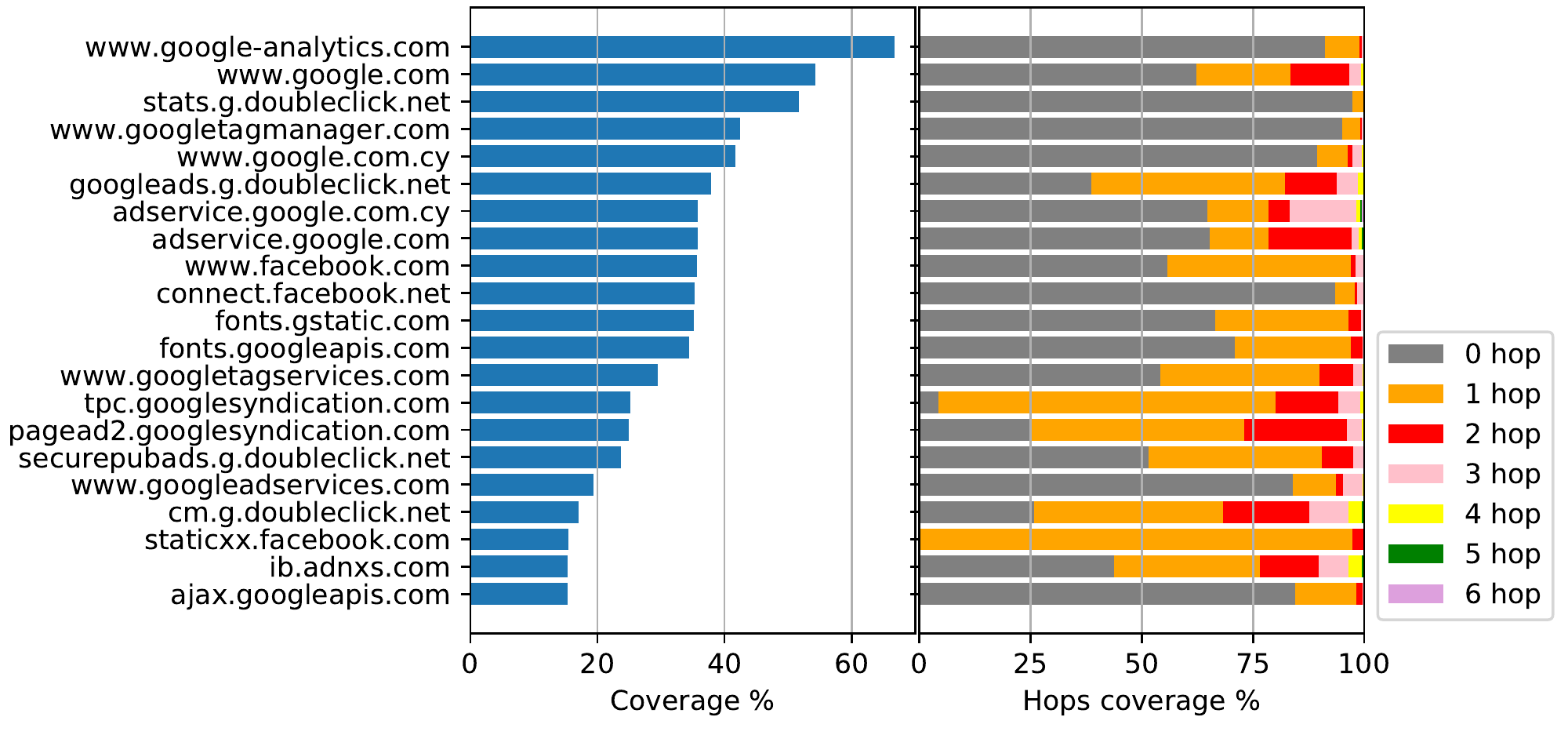} }}
 \caption{The top 20 third party domains coverage percentage (left half) and The percentage (right half) of the 
inclusion level in hops that we detect them in the TopK category.}
\label{fig:top20-trackers-with-hop-stack}

\end{figure}

\begin{figure*}[!bpt]
	\centering
	\subfloat[Health] {{\includegraphics[width=.7\columnwidth]{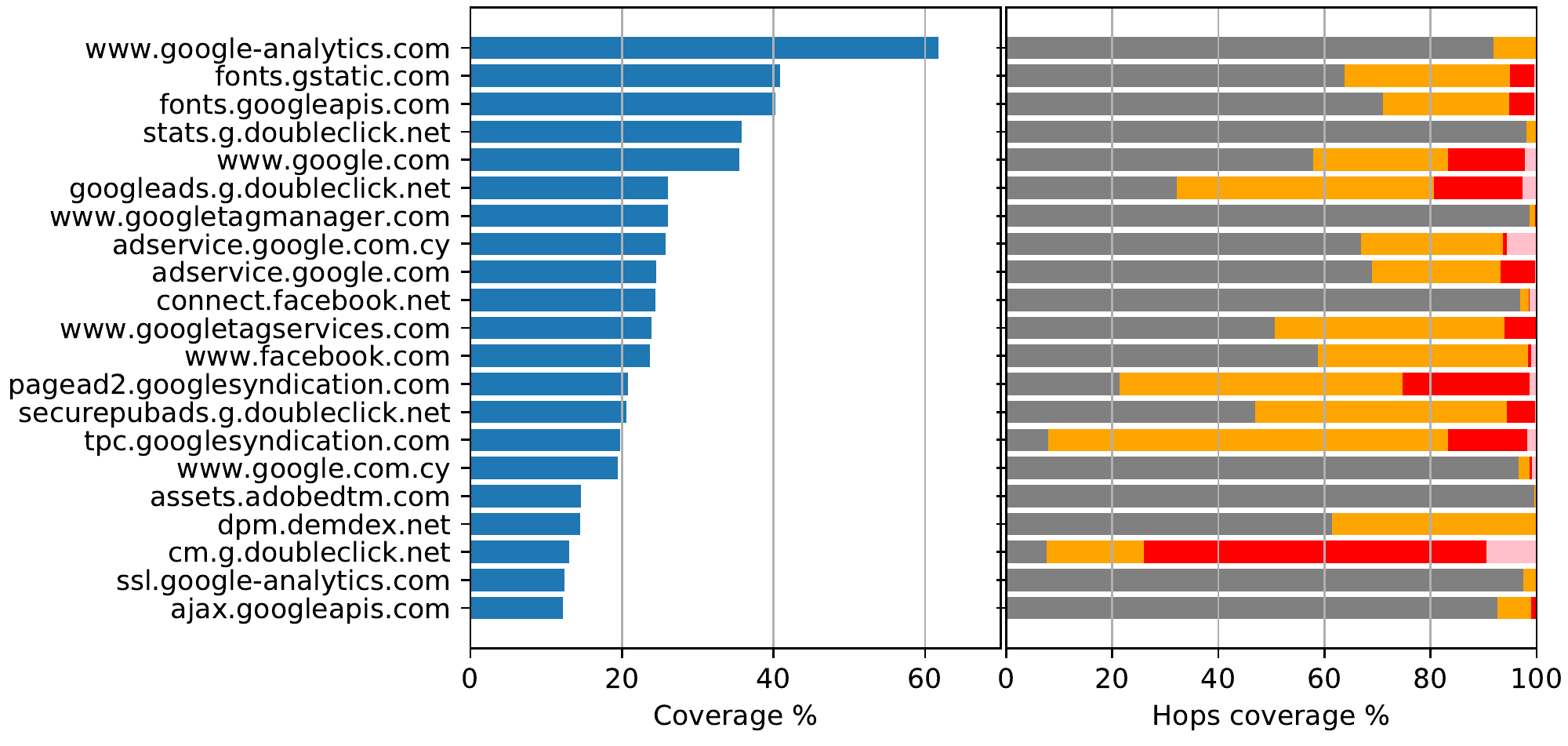} }}
	\subfloat[Ethnicity]{{\includegraphics[width=.7\columnwidth]{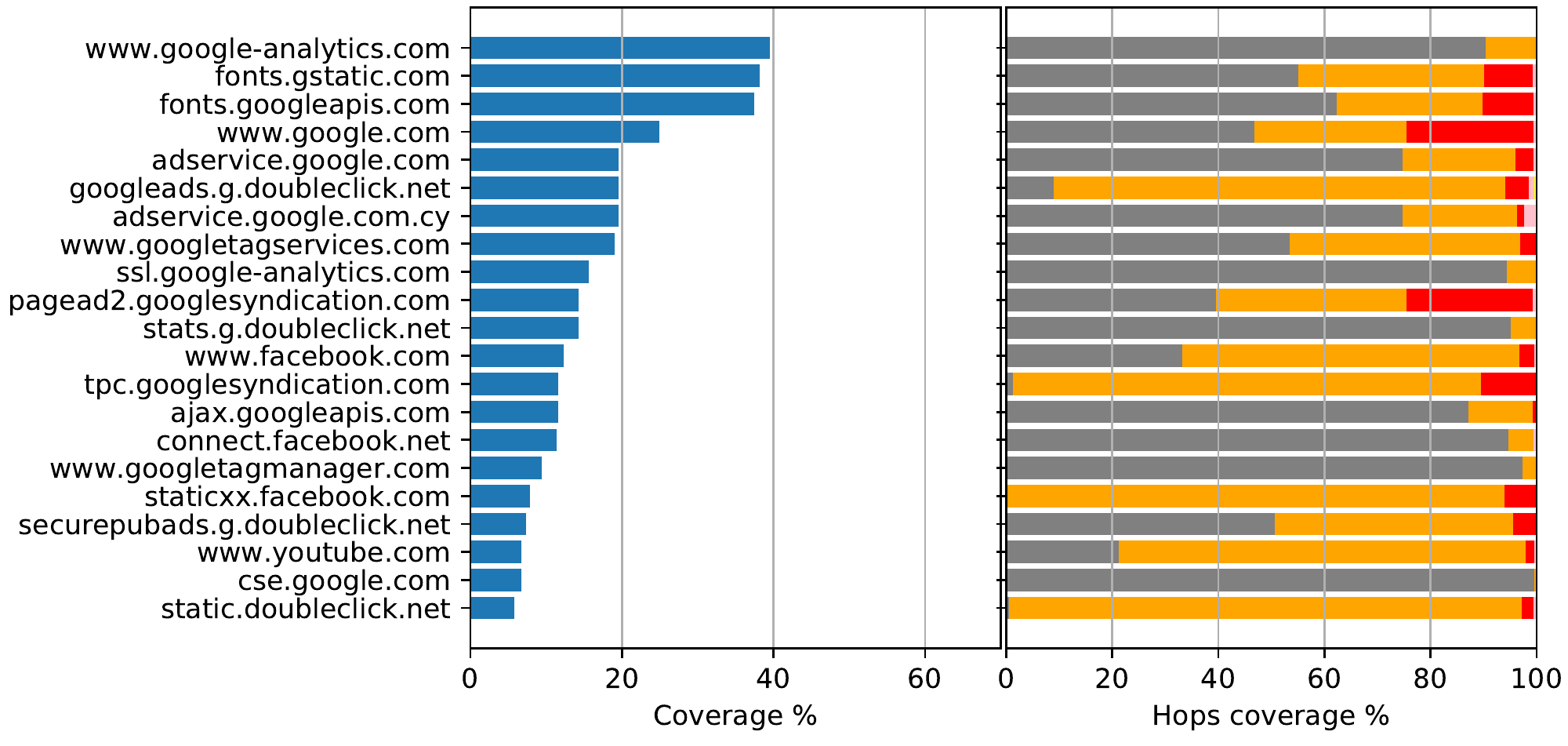} }}
	\subfloat[Religion]{{\includegraphics[width=.7\columnwidth]{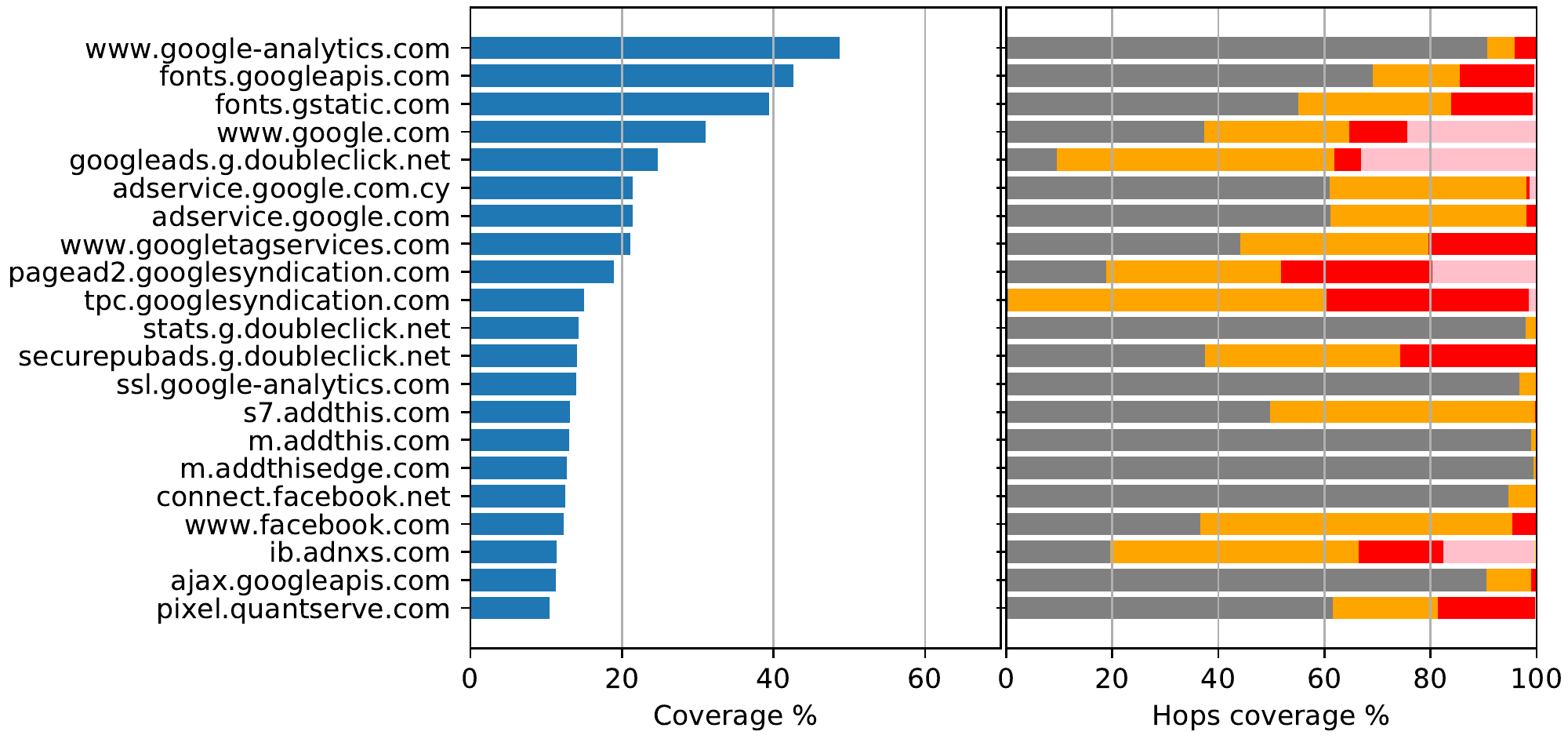} }}\\
	\subfloat[Sexual Orientation]{{\includegraphics[width=.7\columnwidth]{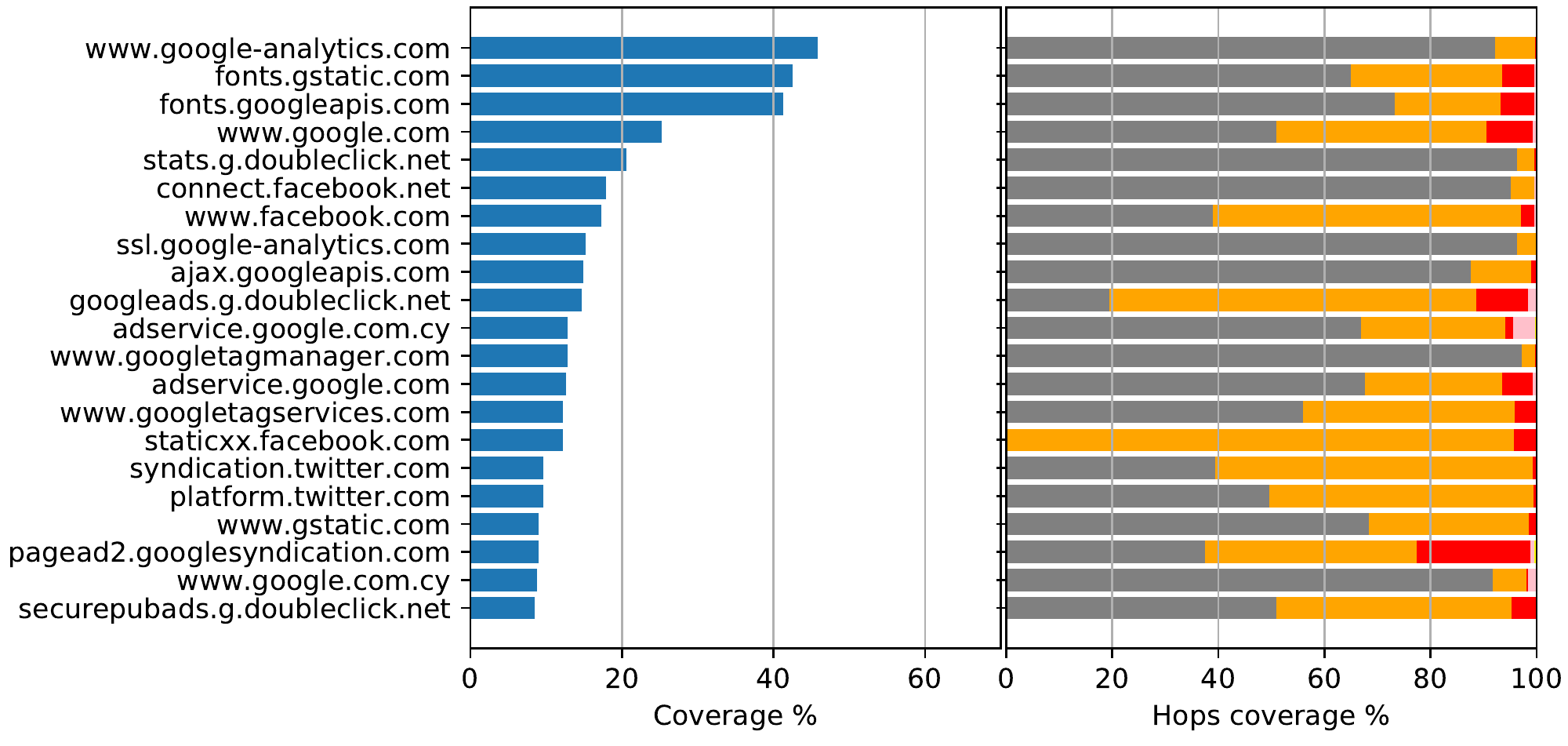} }}
	\subfloat[Political Beliefs]{{\includegraphics[width=.7\columnwidth]{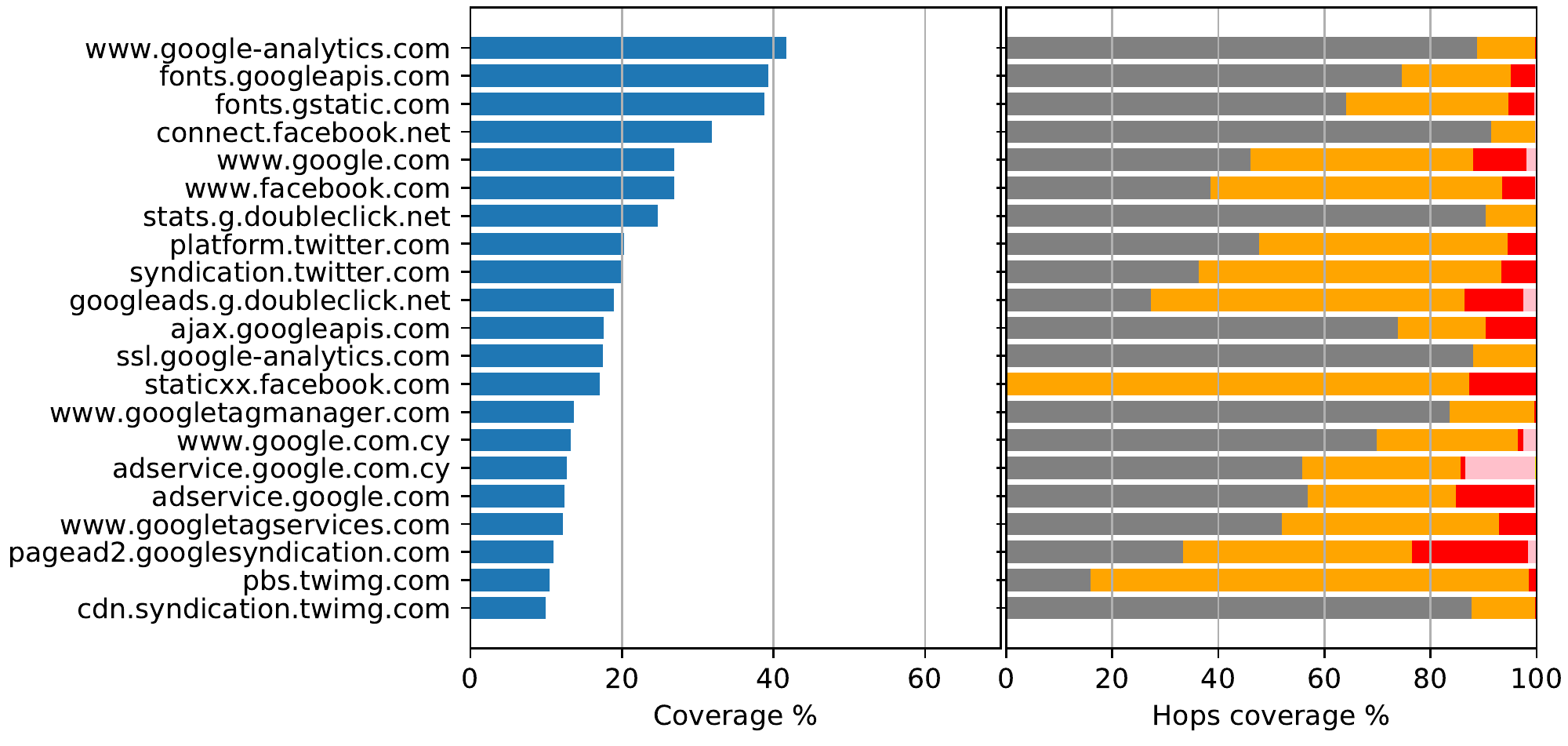} }}
	\subfloat[Porn]{{\includegraphics[width=.7\columnwidth]{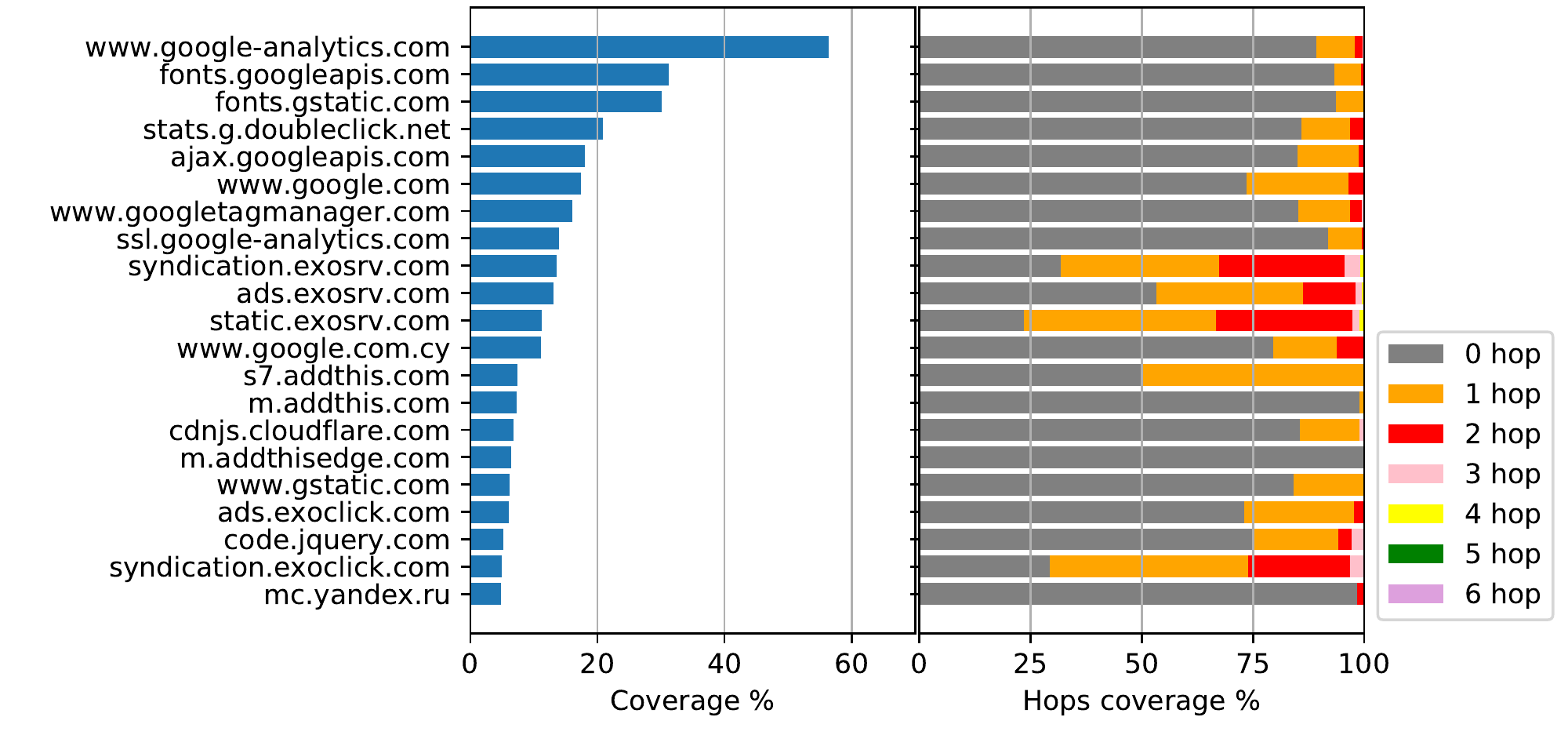} }} 
	\caption{The top 20 third party domains coverage percentage (left half) and The percentage (right half) of the
inclusion level in hops that we detect them in each sensitive category.}    
\label{fig:top-trackers-with-hop-stack-sensitive}
\end{figure*}

\subsection{Are sensitive domains tracked?}

They most certainly are! Table~\ref{table:summary-3rd-parties} shows statistics regarding the number of third party HTTP(S)
requests and domains (full and second level domain) found on TopK and sensitive websites. 
A first observation is that, although both the mean and median number of third party requests per website 
in sensitive websites is significantly less than in TopK, there is a significant number of third party HTTP requests.
The median number ranges from 14 (Ethnicity) up to 40 in (Porn).  
When we turn our attention to third party domains, overall, the median and mean of third party full domains per website in sensitive
websites is lower than this in TopK. Nevertheless, there are third party domains in sensitive websites, where visitors
of such websites would prefer not to be tracked. Focusing on the sensitive categories, we notice that the number of
third parties varies across categories. The median number of third parties in sensitive domains range from as high as 10
for Health related sites, to as low as 5 for Ethnicity related ones (when the corresponding median for TopK sites is
17). We also report the numbers for the second level domain, i.e., TLD+1. We do this as we want to remove bias due to
ephemeral full domains that may be generated by some of the third parties in the corpus of the webpages we study. The
qualitative observations, however, do not change.  The corresponding median numbers when considering TLD+1 are 8 for Health
(highest) and 4 for Ethnicity (lowest), when for TopK the corresponding number is 12. The qualitative observation remain the same also when we consider corporate relationships between domains (\eg domains belonging to the same mother company~\cite{Bashir2016}).
The above results show that although clearly less intense than in TopK popular domains (where we expect to observe a high
number of third party HTTP(S) requests and domains~\cite{Web-Page-Complexity:ToN2013,Website-Complexity}), an alarmingly high number of third parties is present on sensitive domains across
all the considered categories.

\subsection{Who is tracking on sensitive domains?}


We present the names of the first twenty third party domains with the highest
cover in the TopK in Figure~\ref{fig:top20-trackers-with-hop-stack} and in the different sensitive categories that we
consider in Figure~\ref{fig:top-trackers-with-hop-stack-sensitive}. A quick inspection reveals that all the
encountered third parties belong to well known tracking services. We also observe that the top trackers on sensitive
domains are the same ones known to have the highest tracking coverage among all domains on the web. Their coverage,
although slightly lower than in TopK domains, remains impressively high in all the sensitive categories we examined. For
example google-analytics.com has at least 40\% coverage across all sensitive categories, including Porn, where it
is above 60\% and not far from its corresponding coverage across TopK domains. 

The second part of each subplot indicates the ``inclusion'' hop count for each tracker: 0 hops corresponding to direct
inclusion by the first party domain, 1 hop, inclusion by another third party that is itself directly included by the 1st
party, and so on. We see that the majority of trackers are directly included by the 1st party, some are included at 1
hop, few at 2 hops, and a very small percentage at more than 2 hops. A direct 0 hop inclusion effectively means that a
tracking domain knows without doubt that it is present at the corresponding (sensitive) domain. With additional
recursive inclusion, a tracking domain may or may not know who the 1st party domain is, depending on how the inclusion
was done.


In summary, despite the special nature of sensitive domains, and the restrictions put by data protection regulation around them, sensitive domains are intensely tracked by mainstream tracking service who undoubtedly are aware of their presence on such sites.

\begin{table*}[]
\caption{Top 10 third party domains sorted by sensitive category coverage filter based on other category websites coverage \textless q.}
\vspace{-2mm}
\label{table:profile-specialized-trackers}
\resizebox{1.00\textwidth}{!}{%
\begin{tabular}{|l|l|r|r|l|l|l|l|l|l|l|r|r|l|r|r|l|r|r|}
\hline
\multirow{2}{*}{A/A} & \multicolumn{3}{c|}{Health q = 1.0\%}                                                     & \multicolumn{3}{c|}{Ethnicity q = 5.5\%}                                                  & \multicolumn{3}{c|}{Religion q = 2.5\%}                                                   & \multicolumn{3}{c|}{Sexual Orientation q = 3.5\%}                                         & \multicolumn{3}{c|}{Political Beliefs q = 1.0\%}                                            & \multicolumn{3}{c|}{Porn q = 2.0\%}                                                       \\ \cline{2-19} 
                     & \multicolumn{1}{c|}{Domain} & \multicolumn{1}{c|}{Cat. \%} & \multicolumn{1}{c|}{Oth. \%} & \multicolumn{1}{c|}{Domain} & \multicolumn{1}{c|}{Cat. \%} & \multicolumn{1}{c|}{Oth. \%} & \multicolumn{1}{c|}{Domain} & \multicolumn{1}{c|}{Cat. \%} & \multicolumn{1}{c|}{Oth. \%} & \multicolumn{1}{c|}{Domain} & \multicolumn{1}{c|}{Cat. \%} & \multicolumn{1}{c|}{Oth. \%} & \multicolumn{1}{c|}{Domain}   & \multicolumn{1}{c|}{Cat. \%} & \multicolumn{1}{c|}{Oth. \%} & \multicolumn{1}{c|}{Domain} & \multicolumn{1}{c|}{Cat. \%} & \multicolumn{1}{c|}{Oth. \%} \\ \hline
1                    & moatads.com                 & 11.17                        & 0.86                         & ggpht.com                   & 5.40                         & 5.08                         & contextweb.com              & 9.33                         & 1.37                         & maps.googleapis.com         & 4.29                         & 3.21                         & fbsbx.com                     & 6.21                         & 0.14                         & exosrv.com                  & 14.06                        & 0.22                         \\ \hline
2                    & media.net                   & 9.22                         & 0.63                         & 2mdn.net                    & 4.83                         & 3.99                         & sonobi.com                  & 9.13                         & .37                          & yimg.com                    & 4.26                         & 2.01                         & imrworldwide.com              & 4.73                         & 0.89                         & exoclick.com                & 7.68                         & 0.09                         \\ \hline
3                    & ibclick.stream              & 8.70                         & 0.05                         & paypal.com                  & 4.79                         & 0.82                         & districtm.io                & 9.11                         & 0.69                         & consensu.org                & 3.52                         & 3.24                         & nationbuilder.com             & 4.58                         & 0.06                         & nsimg.net                   & 5.07                         & 0.02                         \\ \hline
4                    & webmd.com                   & 8.55                         & 0.04                         & lycos.com                   & 4.54                         & 1.90                         & storage.googleapis.com      & 9.03                         & 0.33                         & gravatar.com                & 3.37                         & 2.34                         & d3n8a8pro7vhmx.cloudfront.net & 4.00                         & 0.04                         & yandex.ru                   & 4.76                         & 1.68                         \\ \hline
5                    & moatpixel.com               & 8.31                         & 0.14                         & lygo.com                    & 4.35                         & 1.81                         & dtyry4ejybx0.cloudfront.net & 9.01                         & 0.00                         & lycos.com                   & 3.02                         & 2.03                         & 3lift.com                     & 3.51                         & 0.65                         & yadro.ru                    & 4.15                         & 0.97                         \\ \hline
6                    & medscape.com                & 7.89                         & 0.03                         & tmdn2015x9.com              & 4.31                         & 1.80                         & secureaddisplay.com         & 9.01                         & 0.00                         & lygo.com                    & 2.81                         & 1.94                         & outbraining.com               & 3.15                         & 0.56                         & tsyndicate.com              & 3.30                         & 0.07                         \\ \hline
7                    & sharethrough.com            & 6.45                         & 0.45                         & px12015x1.com               & 4.31                         & 1.80                         & deployads.com               & 8.96                         & 0.11                         & tmdn2015x9.com              & 2.81                         & 1.93                         & sitescout.com                 & 2.99                         & 0.93                         & highwebmedia.com            & 2.99                         & 0.01                         \\ \hline
8                    & honcode.ch                  & 6.42                         & 0.07                         & spotscenered.info           & 4.31                         & 1.44                         & 33across.com                & 8.91                         & 0.68                         & px12015x1.com               & 2.81                         & 1.93                         & disqus.com                    & 2.59                         & 0.96                         & statcounter.com             & 2.84                         & 1.64                         \\ \hline
9                    & medtargetsystem.com         & 5.89                         & 0.03                         & lexity.com                  & 4.19                         & .36                          & tapad.com                   & 4.05                         & 2.24                         & translate.googleapis.com    & 2.66                         & 1.65                         & unrulymedia.com               & 2.51                         & 0.17                         & air2s.com                   & 2.84                         & 1.64                         \\ \hline
10                   & qualtrics.com               & 5.26                         & 0.36                         & maps.googleapis.com         & 3.64                         & 3.28                         & w55c.net                    & 3.85                         & 1.78                         & sharethis.com               & 2.63                         & 1.74                         & indexww.com                   & 2.32                         & 0.75                         & securedataimages.com        & 2.53                         & 0.01                         \\ \hline
\end{tabular}%
}
\end{table*}

\subsection{Are there specialized trackers for sensitive domains?}

Having seen that the top spots in tracking sensitive domains are occupied by well known trackers, we look further down
the list for less known ones. The reason for doing this goes beyond mere curiosity. As explained in~\cite{Laoutaris18:ConcernsProspects}, mainstream trackers are under intense media, regulatory, and investigative scrutiny and thus would risk a lot if they behaved carelessly, especially with matters relating to sensitive data. On the other hand, large numbers of smaller trackers fly totally under the radar and might, thus, prove to be more dangerous. Next, we first describe briefly how we searched for such trackers, and then proceed to present our findings.  

\subsubsection{Methodology}
In order to examine if there exist ``niche'' third party trackers with a bias towards  sensitive categories, we apply the following methodology. We estimate the coverage of a third party tracker in a sensitive category as the number of websites in which we observe it, divided by the total number of websites in the sensitive category. Then, we exclude third party trackers with high percentage coverage in the rest of websites not belonging to the specific sensitive category that we consider.

In order to control the number of excluded third party trackers we define a maximum coverage threshold \textit{``q''}, such that, if the
percentage coverage of a third party tracker on general websites beyond the particular sensitive category that we consider is above \textit{``q''}, we
exclude the third party tracker from the list of suspected niche trackers for the category. In Table~\ref{table:profile-specialized-trackers} we report the top 10 niche third party trackers found in each category sorted based on their corresponding coverage within the category. 
The value of the \textit{``q''} threshold used is reported next to each category name in the first row of the table.

\eat{
\nlnote{Who is doing that? Has to be brief.}

\todo{We have to update this methodology}

1. Reduce domain names at TLD+1 for both 1st and 3rd party domains (avoid multiple visits to the same 1st party domain).

2. Exclude 3rd party domains found only in 1 first party domain

3. Exclude 3rd party trackers that exists only in Top20k

4. Exclude 3rd parties with maximum coverage in Top20K

5. Keep 3rd parties that have at least one coverage in any sensitive topic above the Top20K coverage.  
I.e. IF MAX(Top20k, Religion, Health, Sexual, ... ) == Top20k THEN skip

\todo{We use a semi-supervised technique: We may want to use a p-q filtering, i.e., appear more than p\% in websites in a specific category and less than q\% in the
non-sensitive/top k sites.}

\todo{Sensitivity q value} In Figure~\ref{fig:p-q-sensitivity}
}

\subsubsection{Findings}

Table~\ref{table:profile-specialized-trackers} depicts the names and the coverage of the top-10 niche trackers for each sensitive category for different values of $q$. We see that such trackers can achieve up to 14\% coverage within a sensitive category. This is much lower than the coverage achieved by mainstream trackers across all domains, which is probably expected given that such trackers serve niche markets and are, themselves, much smaller companies.


We manually looked up the discovered trackers and verified that most of them offer tracking services, whereas some declared clearly on their web-sites that they indeed specialize on the categories that we found them to specialize on.
For example, medtargetsystem.com, owner of domain dmdconnects.com (position 9 in category Health of Table~\ref{table:profile-specialized-trackers}) describes clearly its services for medical professionals and web-sites. At position 76 of Health (not depicted on the table) we find ehealthcaresolutions.com, describing itself as a ``\emph{a unique marketing platform that specializes in connecting niche audiences with pharmaceutical and healthcare brands}''. At position 89, tapnative.com, offers to ``\emph{promote your content to millions of health-conscious consumers and healthcare professionals exclusively within premium health and wellness websites and around health-related content.}''

Going over to Politics, on position 3 of Table~\ref{table:profile-specialized-trackers} we find nationbuilder.com stating on their web-site: ``\emph{We need more leaders. And better ones. Wherever you are on your path to leadership, NationBuilder Cities will equip you with valuable tools in real life, starting with scheduled events to help you build community, share stories, and gain the skills and training you need to lead in this era.}''

%

At position 96 of Sexual Orientation we find codeamber.org describing its services as follows: ``\emph{Who would want a Background Check? Anyone who wants to know more about someone new in their life may well want to run a deep background check on that individual. Whether it's a new boy or girl friend, a new neighbor or someone new in your, or your child's life who may be or become significant. Don't take a chance. Get a complete background report on them today.}''.

Finally, on position 11 of Porn we encounter JuicyAds, self-described as ``\emph{the sexy advertising network}''.



%% file: sections/communication.tex
\section{Communication Between Trackers}\label{sec:communication-trackers}

In this section, we look at the communication patterns between trackers operating on sensitive domains. We are
particularly interested in investigating whether there are signs of exchange of information between mainstream third party trackers that may hold
Personally Identifiable Information (PII) for users (email address, first and last name, \etc), and niche third party trackers, that
typically don't have PII, but may have seen users in a series of sensitive domains. 

\begin{table*}[!bpt]
\caption{Statistics on cookie synchronization.}
\label{table:trackers-in-sensitive-domains}
\resizebox{2\columnwidth}{!}{%
\begin{tabular}{|c|c|c|c|c|c|c|c|c|c|c|}
\hline
Category & \#Websites & \#Domains & \#Websites & \%Websites & \#HTTP   & \#HTTP Req. & \%HTTP Req. &\#Unique & \#CSync  & \%CSync \\ 
		 & 			  & 		  & with Csync & with Csync & Requests & with CSync  & with CSync  & Pairs   & niche pairs &niche pairs\\ \hline \hline
Top k & 7,115 & 7,115 & 1,331 & 18.71 & 870,954 & 13,739 & 1.58 & 2,460 & 145 & 5.9 \\ \hline
Health & 7,144 & 3,989 & 1,072 & 26.87 & 432,332 & 5,953 & 1.38 & 327 & 42 & 12.8 \\ \hline
Ethnicity & 3,126 & 2,299 & 90 & 3.91 & 93,744 & 619 & 0.66 & 167 & 7 & 4.2 \\ \hline
Religion & 10,134 & 7,018 & 588 & 8.38 & 464,192 & 8,675 & 1.87 & 296 & 17 & 5.7 \\ \hline
Sexual Orientation & 3,373 & 2,787 & 143 & 5.13 & 141,729 & 1,136 & 0.8 & 192 & 16 & 8.3 \\ \hline
Political Beliefs & 5,516 & 3,874 & 372 & 9.60 & 379,887 & 5,639 & 1.48 & 308 & 36 & 11.7 \\ \hline
Porn & 1,301 & 1,293 & 5 & 0.39 & 92,104 & 107 & 0.12 & 36 & 0 & 0.0\\ \hline
All Sensitive  & 30,594 & 20,688 & 2,270 & 10.97 & 1,603,988 & 22,129 & 1.38 & 843 & 67 & 7.0\\ \hline
Overall & 37,709 & 27,697 & 3,601 & 13.0 & 2,474,942 & 35,868 & 1.45 & 3,006 & 162 & 5.4 \\ \hline
\end{tabular}
}
\end{table*}

\subsection{Cookies Synchronization}\label{sec:cookie-synchronization}


Cookie synchronization~\cite{Bashir2016,cookies-sync2019,castelluccia_2014}, or simply CSync, is a well known technique used by  Ad
and Tracking entities to synchronize pseudonymous user IDs that the different entities assign to users. Due to the same
origin policy~\cite{SOP} that prevents two third-party domains to directly exchange information between each other,
third-party domains need to use alternative methods to exchange information. To bypass the same origin restriction, two
third-party domains can synchronize their cookies by passing them as arguments in the URL of an HTTP(S) GET request. The
HTTP(S) request is usually towards a small image 1$\times$1 pixel, hosted under the second third-party domain. The cookie
of the first third-party domain is embedded in the request URL as an argument and the cookie of the second third-party
domain (for the same user) is embedded in the request header. Note that the default web browser behavior is to include
existing cookies belonging to the visited domain in the request header. Upon successful cookie synchronization, the two
third-party domains can exchange tracking information in the background related to the same user. Thus, they enrich
their data about their existing users and increase their visibility to other first-party domains in which they do not have
presence but their partners do. For more details related to CSync see~\cite{cookies-sync2019}.



\subsection{Detecting Cookie Synchronization}\label{sec:detection-cookes-sync}


To detect CSync, as a first step, we exclude all requests between third-party domains that do not
include any arguments in the URL. Then, we filter the URL using a list of keywords that we empirically build related to
CSync activities. Some keyword examples are, \textit{``usercookie'', ``external\_user\_id'', ``usermatch'',
``async\_usersync''}, \textit{etc.} In total, we have 62 keyword. We also exclude URLs with obfuscated arguments since
we cannot examine if they include any user related information or cookies. We also assume that all sub-domains belonging
to the same top level domain plus one (TLD+1), \textit{i.e.,} ``pagead2.googlesyndication.com'' and
``tpc.googlesyndication.com'', can access cookies belonging to the TLD+1 ``googlesyndication.com'', thus, by default
they can share information for the same user.

More advance CSync detection techniques~\cite{Bashir2016,cookies-sync2019} can also be incorporated in our methodology, nevertheless, comparing our results with those reported in~\cite{cookies-sync2019} we observe that the percentage of additional CSync requests is relatively small. Note that since we collect data with a crawler that has full control of the web browser we can also analyse HTTPS requests.

\eat{
\todo{- How do we identify cookies sync when HTTPS or encryption of IDs is used? Discuss e.g., the case of DoubleClick?}

\todo{- How do we distinguish between CSync and RTB?}

\todo{-do we discuss cookie-less synchronization? Is it possible to track the user, even if he/she deletes the
cookie?\cite{CCS2013-Web-never-forgets}.}

- papers~\cite{Bashir2016,cookies-sync2019} discuss the above issues and we have to make sure we use the state of the
  art techniques for CSync.

\todo{Do we see anything in the plain text regarding the data that is exchanged, e.g., sensitive information?
\cite{tracking-trackers}}
}

\subsection{Statistics on Cookie Synchronization}\label{cookies-sync-stats}
Next, we turn our attention towards ``who talks to whom?''. In Table~\ref{table:trackers-in-sensitive-domains} we report the number of websites, across categories including, the TopK and the five sensitive categories (first column), where we identified cookie synchronization in our study. The second and third column depicts the number of websites and the number of domains in each category, respectively. The forth column the number of websites we detect atleast one CSync instance, and fifth column the corresponding percentage.  

When we turn our attention to the percentage of
websites with synchronization (fifth column), we notice that this varies significantly across different categories. 
Around 19\% of popular websites (TopK) host third parties that exchange information using cookie synchronization.
In websites belonging to the sensitive categories this percentage is significantly lower, typically below 10\%, with the exception of Health, in which 27\% of the websites we examined host third
parties that participate in cookie synchronization. On the other extreme, in Porn, only 0.4\% of websites host third parties that participate in cookie synchronization. 

With respect to HTTP(S) requests involved in cookie synchronization (column six, seven and eight). For the popular TopK websites, the percentage of HTTP(S) requests that are used for cookies synchronization is 1.58\%. Our results agree
with other recent studies that studied cookie synchronization, e.g., the one in~\cite{cookies-sync2019}, where the
authors reported that 1.47\% of the HTTP requests generated from 850 real mobile users over one year are used for cookie synchronization.
Again, comparing the percentages of the sensitive categories with the TopK, we observer much lower percentages is sensitive categories.

Column nine shows the absolute number of CSync pairs that we detect in each category. We observe that in TopK we detect 2,460 pairs followed by Health with 327 and Religion with 296 pairs. The rest of the categories are below 200 pairs.

Next, in columns ten and eleven we show the absolute number (column ten) of CSync pairs between the niche tracking domains reported in Table~\ref{table:profile-specialized-trackers} and the percentage of CSync pairs in each category belonging to those trackers.
We observe that Health and Political Beliefs have 12.8\% and 11.7\% of CSync pairs involved with niche trackers of each category, respectively.

In summary, cookie synchronization is less common in sensitive websites (except Health), than in popular ones.  Porn is a
sensitive category where the cookie synchronization is rare. Nevertheless, there is exchange of information between mainstream trackers (that have PII) and niche trackers that don't have PII but have seen people on sensitive domains. Given that the latter fly largely under the radar, and may be prone to more risky exploitation plans for the data they collect, we find such exchange of information worrisome. 

\subsection{Limitations}

We believe that
most of the cookie synchronizations
between trackers are visible in our study. We, however, can not exclude other types of communication between 
trackers, e.g., backend server-to-server communication based on private contracts, that are not visible within the browser (In our case, our crawler). 
Moreover, trackers are added in the webpages and new contracts for exchange of
information between them are signed at a regular basis. Thus, in Table~\ref{table:trackers-in-sensitive-domains}, we report lower bounds based on a snapshot of the cookie synchronization activity, as we observe in our study.

%% file: sections/related.tex
\section{Related Work}\label{sec:related-work}

A substantial amount of work has touched upon web tracking across different platforms, such as, desktop
computers~\cite{Balebako_2012, BashirIMC2018, Fruchter_2015, WebTracking-over-years, Pujol_2015,
openWPM-englehardt2016census, Walls_2015}, mobile phones and tablets~\cite{Mobile-Apps-NDSS2018, Binns2018, ReCon,
Leung_2016} or mixed platforms (Mobile apps and Web interfaces)~\cite{Gervais2017, Leung_2016, Mobile-Apps-NDSS2018}.

More specifically, Lerner et. al.,~\cite{WebTracking-over-years} study the evolution of web tracking over time (1996-2016). 
Steven et. al.,~\cite{openWPM-englehardt2016census} measure the extend of web tracking in the top 1 Million websites. 
The authors in~\cite{BashirIMC2018, Ads-vs-Regular-Contents} study the privacy impact of web tracking and web ads. 
In a similar path, the authors in ~\cite{Balebako_2012, Pujol_2015, Walls_2015, Gervais2017} study the impact on web tracking when using AdBlocking tools. 
With respect to personal data leakage, the authors in~\cite{ReCon, Starov_2016} study the leakage of Personal Identifiable Information (PII). 
Another aspect of the web tracking  recently studied is the geographic location of tracking servers~\cite{Fruchter_2015, Mobile-Apps-NDSS2018, Iordanou:GDPR-IMC2018}.

Few studies include results on tracking of sensitive domains~\cite{Mayer_2012,WPES12,Mobile-Apps-NDSS2018,CoNEXT2015-Following,Iordanou:GDPR-IMC2018}. In these works the study of sensitive domains is just a small part of a longer study. Some works looking on tracking and targeting of minors and COPPA related violations also fall in the area of sensitive domain tracking~\cite{Reyes2018}. Compared with all these works our results have greater scale (\# sensitive domains examined, \# sensitive categories), greater generality (our methodology can easily be adapted to monitoring arbitrary sensitive categories), and ask questions that have not been asked before (``are tracking domains aware that they operate on sensitive domains?'', ``are there specialized trackers?'', ``do specialized trackers talk to mainstream ones?'').

Text and Web domain classification includes a large body of scientific literature (see~\cite{Kowsari2019} for a recent
survey) as well as several commercial services, both standalone~\cite{alexaTop, similarweb}, or as parts of the campaign
planers of online advertising services~\cite{google_display_planner}. Such services are payment based and opaque as far
as their internals are concerned. Also, most of them avoid including very sensitive terms since this would constitute an
obvious violation of data protection rules (referring to campaign planners).  Web-domain and text based classification
are huge areas upon which we draw tools like TF-IDF~\cite{Salton1988} and BoW~\cite{Ko2012} for feature engineering,
Na\"ive Bayes algorithm~\cite{Adetunji2018} for classification, \etc Our contribution in the area is more on terms of
how we combine things together rather on fundamental tools. Curlie.org~\cite{curlie} is an ideal taxonomy for finding
large lists of sensitive domains really fast, even manually, due to the hierarchical organization of the information.
By combining the above with the previously mentioned tools it is easy to produce accurate classifiers for
automatically identifying large sensitive domains among arbitrary lists of domains on the open web. 


To the best of our knowledge, the proposed methodology in this work is the first generic methodology that allows a quick and easy way to identify a large number of websites belonging to different categories, with a minimal effort to bootstrap and use.
The simplicity and generality of the proposed methodology makes it suitable for a plethora of applications in the
measurement community to measure and analyse websites based on different categories or similarities.  This generality
can be achieved during the training phase of the classifier by selecting training samples related to the topic of
interest or any other common characteristics of the websites depending on the goals of each study.
In this paper, we showcase just a simple example, how to use it in order to monitor web tracking on sensitive websites as defined by GDPR in Europe. 

%% file: sections/conclusion.tex
\section{Conclusion}\label{sec:conclusion}

Given the recent intense debates around data protection, we've been surprised to find so many tracking services operating
on carefully identified sensitive domains. Of course, one may arrive to such domains accidentally by, say, clicking on
the wrong link, or because one's device has been hacked. Most visits, however, are intentional, and, therefore, reveal
sensitive information about the visitor. Our work has shown that in the majority of cases, it is the owners of such sites
that intentionally include tracking code in order to participate in advertising revenue sharing programs of large and
small online advertising companies. Such companies, given the direct inclusion of their tracking code by the site owner,
are clearly aware of their presence on such sites. To avoid being there, they would have to refuse admission to their
revenue sharing programs to sites that handle sensitive topics. This would require some manual filtering but, as our
work has demonstrated, most of the effort can be automated. On the other hand, operators of such sites need to receive
revenue to keep their sites going, and this implies participation in ad revenue sharing platforms. Perhaps, there should
be a special way to handle tracking and advertising for sensitive domains. We intent to examine such ideas as part of
our future work.